\documentclass[twocolumn]{aastex7}

\begin{document}

\renewcommand{\topfraction}{1.0}
\renewcommand{\bottomfraction}{1.0}
\renewcommand{\textfraction}{0.0}

\newcommand{\kms}{km~s$^{-1}$\,}
\newcommand{\masyr}{mas~yr$^{-1}$\,}
\newcommand{\msun}{$M_\odot$\,}

\title{Orbits of Twelve   Multiple  Stars}

\author{Andrei Tokovinin}

\affiliation{Cerro Tololo Inter-American Observatory, NSF's NOIRLab
Casilla 603, La Serena, Chile}
\email{andrei.tokovinin@noirlab.edu}

\begin{abstract}
Inner  and outer  orbits in  twelve hierarchical  stellar systems  are
determined using  high-resolution speckle imaging,  radial velocities,
or  both.  Masses  and fluxes  of the  components are  estimated.  The
Hipparcos numbers  of the  main stars are  7111, 12912,  17895, 20375,
42424, 68717, 77439, 79076, 90253, 97922, and 102855; the faint triple
WDS J10367+1522 has  no HIP number. Four systems are  quadruple of 3+1
hierarchy,  the rest  are triple.   Two triples  with low-mass  M-type
components are approximately planar,  with moderately eccentric orbits
and near-unit mass ratios. The shortest inner period of 0.27d is found
in  the  newly identified  contact  eclipsing  pair belonging  to  the
misaligned  quadruple  HIP~97922.   The   compact  system  HIP  102855
(periods 15.4 and  129 days) identified by Gaia is  confirmed here and
has additional companion  at 6\arcsec. This work  contributes new data
for the study  of diverse architectures of stellar  hierarchies in the
field.
\end{abstract}





\section{Introduction}
\label{sec:intro}

This  paper reports  on the  orbital parameters  and masses  in several
hierarchical stellar systems, deduced  from original observations. It
is an  incremental contribution to  the collection of data  on stellar
hierarchies     in     the      Multiple     Star     Catalog,     MSC
\citep{MSC}.\footnote{The  latest   MSC  versions  are   available  at
  \url{https://ww.ctio.noirlab.edu/~atokovin/stars}             and
  \url{http://vizier.u-strasbg.fr/viz-bin/VizieR-4?-source=J/ApJS/235/6}. }
  
Bound systems of three or more stars are relics of star formation, and
their  architecture  (masses,  periods, eccentricities,  mutual  orbit
orientation) contains valuable information on  the origin of stars and
planets,  both  statistically  and individually.   The  statistics  of
stellar  hierarchies, poorly  known at  present, are  also a  starting
point  for modeling  their  evolution which  produces unusual  stellar
remnants  and   high-energy  events   \citep{Toonen2020}.   Simplistic
attempts  to model  the population  of triple  systems as  independent
combinations  of two  binaries drawn  from a  common distribution  and
filtered  by  dynamical   stability  \citep[e.g.][]{Fabrycky2007}  are
caused by the lack of actual data.  Indeed, owing to the vast range of
periods, a  relatively complete  volume-limited sample  of hierarchies
exists only for the well-studied nearest solar-type stars \citep{R10},
and  it is  necessarily small  (56 systems).   Extending the  distance
limit  to 67\,pc  reveals  a  patchy and  incomplete  coverage of  the
parameter space \citep{FG67b}.

A long-term program of monitoring radial velocities (RVs) of late-type
stars  in hierarchical  systems  (mostly within  67\,pc)  in order  to
determine unknown inner periods has been started by the author in 2015;
it   contributed  $\sim$100   spectroscopic   orbits  and   discovered
additional, previously unsuspected subsystems \citep{chiron10}.  After
termination  of the  project,  some stars  still lacked  spectroscopic
orbits, prompting continued observation.   Results on one such system,
HIP~68717,  are published  here.   This work  also  revealed that  the
latest Gaia catalog of non-single  stars, NSS \citep{NSS}, covers only
a third  of our sample.   Furthermore, not all  orbits in the  NSS are
correct  \citep{Holl2023,chiron11}.   This is  understandable  because
signals from multiple systems are often non-trivial and unsuitable for
automatic pipelines such as Gaia.

The  situation is  even less  satisfactory  in the  period range  from
decades  to millenia,  where the  RV monitoring  lacks time  coverage,
while  the all-sky  Gaia data  lack both  spatial resolution  and time
coverage.   High-resolution   imaging  of  binary  stars   by  speckle
interferometry at the the 4.1 m SOAR (Southern Astrophysical Research)
telescope   produced   hundreds   of  serendipitous   discoveries   of
hierarchical systems  by resolving inner subsystems  in known binaries
or  by detecting  their tertiary  companions.  For  nearby stars,  the
diffraction-limited resolution  of 30\,mas gives access  to relatively
short orbital periods,  allowing calculation of orbits:  the inner one
and,  when  historic  data  are  available, the  outer  one  as  well.
Accumulation  of  speckle data  enables  a  detailed analysis  of  the
system's  architecture  \cite[see][and  references  therein]{Tok2025}.
Measurements  of  the  mutual  orbit  inclination  are  of  particular
interest  for studying  the origin  of stellar  hierarchies and  their
dynamical evolution \citep{Mult2021}.  This paper continues the series
based on the  SOAR speckle data. Thus, it combines  the speckle and RV
work.

The basic  data on the  12 hierarchies  studied here are  assembled in
Table~\ref{tab:objects}. The systems are  identified by the Washington
Double Stars (WDS) codes based  on the J2000 positions \citep{WDS} and
by numbers  of their  components in  the Hipparcos  and HD  catalogs. The
spectral  types and  photometry  are copied  from  Simbad, the  proper
motions (PMs) and parallaxes are, mostly, from the Gaia data release 3
(GDR3) \citep{Gaia3}, and most RVs are from this work.  Each hierarchy
is  discussed  individually in  the  following  sections. To  give  an
overall  view of  their parameters,  Table~\ref{tab:masses} lists  the
number  of known  components  $N$  and the  hierarchy  in the  bracket
notation, where for  each pair the primary and  secondary masses $m_1$
and $m_2$  and the orbital period  $P$ are listed.  There
are 8  triples and 4  quadruples. All  quadruples have a  3-tier (3+1)
hierarchy  (a close  inner pair,  an intermediate  pair, and  an outer
companion).  The  orbital periods  range from  0.27 days  to $\sim$200
kyr.

\begin{deluxetable*}{c c rr   l cc rr r c }
\tabletypesize{\scriptsize}     
\tablecaption{Basic Parameters of Observed Multiple Systems
\label{tab:objects} }  
\tablewidth{0pt}                                   
\tablehead{                                                                     
\colhead{WDS} & 
\colhead{Comp.} &
\colhead{HIP} & 
\colhead{HD} & 
\colhead{Spectral} & 
\colhead{$V$} & 
\colhead{$V-K_s$} & 
\colhead{$\mu^*_\alpha$} & 
\colhead{$\mu_\delta$} & 
\colhead{RV} & 
\colhead{$\varpi$\tablenotemark{a}} \\
\colhead{(J2000)} & 
 & &   &  
\colhead{Type} & 
\colhead{(mag)} &
\colhead{(mag)} &
\multicolumn{2}{c}{ (mas yr$^{-1}$)} &
\colhead{(km s$^{-1}$)} &
\colhead{(mas)} 
}
\startdata
01316$-$5322     & AB & 7111   & 9438   & F5V    & 7.87 & 1.30  & 86     & 29   & 29.0     & 9.72: DR3 \\
                 & C  & \ldots & \ldots & \ldots & 11.20 & 2.03 & 81     & 30   & 31.2     & 9.68 DR3 \\  
02460$-$0457     & AB & 12912  & 17251  & F3V    & 7.52  & 1.14 & 56     & 7    & \ldots   & 10.69 DR3 \\
03496$-$0220     & AB & 17895  & 24031  & F8V    & 7.23  & 1.48 & $-$59  & $-$41 &  6.0    & 19.63 HIP \\
04218$-$2146     & A  & 20375  & 27723  & G0V    & 7.54  & 1.28 & 86*     & $-$11* & $-$3.7  & 18.99 DR3 \\
                 & B  & \ldots & \ldots & \ldots & 20.24 & 7.42 & 85     & $-$12  & \ldots & 18.84 DR3 \\
08391$-$5557     & ABC & 42424 & 74045  & F3V    & 7.46  & 1.12 & $-$16  &   6   &  26.8:  & 6.22 DR3 \\ 
10367+1522       & AB & \ldots & \ldots & M4.0V  & 13.31 & 5.41 & 110    & $-$78 & \ldots  & 50.0 DR3 \\
14040$-$4437     & AB & 68717  & 122613 & G1V    & 8.28 & 1.86  & $-$9*  & $-$16* & $-$3.3  & 11.76 HIP  \\
                 &  C & \ldots & \ldots & \ldots & 12.43 & 3.37 & $-$13  & $-$20 & $-$1.3   & 12.56 DR3 \\
15474$-$1054     & AB & 77349  & \ldots & M2.5V  & 11.28 & 4.54 & $-$309 & $-370$ & 3.0     & 62.52 DR2 \\
16161$-$3037     & AB & 79706  & 146177 & F0     & 7.74  & 1.12 & $-$26  & $-$21  & \ldots  & 6.20 DR3 \\ 
18250$-$0135     & AB & 90253  & 169493 & F2V    & 6.19  & 1.05 &   5    & $-$3  & $-$12.1  & 7.49 DR2 \\ 
19540+1518       &  A & 97922  & 188328 & F8III  & 7.20 & 1.57  & 0      & $-$12 & 3.9      & 11.78 DR3 \\
                 &  B & \ldots & \ldots & \ldots & 8.65 & \ldots & $-$18 & $-$15 & \ldots   & 11.47 DR3 \\
20503$-$7502     &  A & 102855 & 197324 & F7V    & 8.42 & 1.21  & 22     & $-$64 & 15.1     & 8.29 DR3N \\
                 &  B & \ldots & \ldots & \ldots & 16.63 & 4.63 & 23     & $-$67 & \ldots   & 8.12 DR3 \\ 
\enddata
\tablenotetext{}{Proper  motions  and  parallaxes are  from  Gaia  DR3
  \citep{Gaia3}  or Hipparcos  \citep{HIP2}.   Colons mark  parallaxes
  biased by subsystems, asterisks mark PMs from \citet{Brandt2021}.  }
\end{deluxetable*}

\begin{deluxetable*}{c c  c c c }
\tabletypesize{\scriptsize}     
\tablecaption{Masses, Periods, and Hierarchy
\label{tab:masses} }  
\tablewidth{0pt}                                   
\tablehead{                                                                     
\colhead{WDS} & 
\colhead{HIP} & 
\colhead{$N$} & 
\colhead{Hierarchy}  &
\colhead{Masses and Periods}  \\
\colhead{(J2000)} &  &  & &
\colhead{ $(m_1, m_2; P)$ }
}
\startdata
01316$-$5322  & 7111   &  4 & (A,(Ba,Bb)),C & ((1.37,(1.23,0.92; 4.9yr); 285yr),0.83; $\sim$126kyr) \\ 
02460$-$0457  & 12912  &  3 & A,(Ba,Bb)    & (1.41,(1.1,0.58; 38yr); 1kyr)  \\
03496$-$0220  & 17895  &  3 & A,(Ba,Bb)    & (1.14,(0.98,0.84; 250d); 47yr) \\
04218$-$2146  & 20375  &  3 & (Aa,Ab),B    & ((1.20,0.44; 12.4yr); $\sim$194kyr) \\
08391$-$5557  & 42424  &  3 & A,(B,C)      & (1.92,(1.1481.08; 23.5yr); 1.1kyr) \\
10367+1522    & \ldots &  3 & A,(B,C)      & (0.29,(0.17,0.17; 8.6yr); 120yr) \\
14040$-$4437  & 68717  &  4/5 & ((Aa,Ab),B),(C,?) & (((0.99,0.93; 24d),1.04; 414yr),(0.69+?); $\sim$8kyr) \\
15474$-$1054  & 77349  &  3 & A,(Ba,Bb)    & (0.41,(0.23,0.21; 133d); 8.3yr) \\
16161$-$3037  & 79706  &  3 & (Aa,Ab),B    & ((1.67,1.21; 7.5yr),1.34; 167yr) \\
18250$-$0135  & 90253  &  3 & (Aa,Ab),B    & ((2.28,1.59; 11.75yr),1.96; 386yr) \\
19540+1518    & 97922  &  4 & (Aa,(Ab1,Ab2)),B & ((1.34,(1.00,0.30; 0.27d); 68yr),1.12; 3.1kyr) \\
20503$-$7502  & 102855 &  4 & ((Aa1,Aa2),Ab),B & (((1.31,1.06; 15.4d),0.16; 129d),0.38; $\sim$11kyr) \\
\enddata
\end{deluxetable*}

The  observational   data  and  methods  are   briefly  introduced  in
Section~\ref{sec:data}.   Sections \ref{sec:7111}  to \ref{sec:102855}
are devoted to individual  systems, Section~\ref{sec:sum} contains the
summary.

\section{Data and Methods}
\label{sec:data}

The input  data and their  interpretation are similar to  the previous
papers of  this series.   Below they are  outlined very  briefly, with
details in the references.   The availability of position measurements
and RVs  and their time  coverage differ in each  case.  Additionally,
astrometry and photometry  are used to characterize  each hierarchy as
completely as possible. For this  reason, the discussion of individual
systems  below might  appear difficult  to grasp  and overloaded  with
numbers such as mass estimates.

\subsection{Speckle Interferometry}
\label{sec:speckle}

This work is based on the  long-term monitoring of binary and multiple
stars with high angular resolution using speckle interferometry at the
the  4.1   m  SOAR  telescope   located  in  Chile.    The  instrument
(high-resolution camera), the data processing, and the performance are
covered in \citet{TMH10,HRCam}.  The latest series of measurements and
references to prior observations can  be found in \citet{Tok2024}. The
cubes of short-exposure images are recorded in the $y$ (543/22\,nm) or
$I$ (824/170\,nm)  filters. The pixel scale  (15\,mas) and orientation
are calibrated using  a set of binaries with separations  on the order
of 1\arcsec  and accurately modeled motions.   Typical external errors
of 2\,mas  and less are achieved,  resulting in the most  accurate and
internally consistent set of speckle  data available to date.  This is
essential for  detecting small perturbations (wobble)  caused by inner
subsystems.  Two   triples  studied  here  were   originally  used  as
calibrators  until   the  wobble   was  revealed.    The  measurements
(separation, position angle, and magnitude difference) are obtained by
fitting models of binary or triple  point sources to the speckle power
spectra,  using  single  reference  stars where  necessary  to  reduce
instrumental signatures.

\subsection{Radial Velocities}
\label{sec:RV}

High-resolution  optical spectra  were taken  with the  CHIRON echelle
spectrometer  at  the   1.5  m  telescope  located   at  Cerro  Tololo
\citep{CHIRON} and operated  by the SMARTS consortium.\footnote{SMARTS
  stands for  Small and  Moderate Aperture Research  Telescope System,
  \url{https://www.astro.gsu.edu/~thenry/SMARTS/}} The  spectra with a
resolution  of  80,000  were  processed  by  the  instrument  pipeline
\citep{Paredes2021} and  cross-correlated with a binary  mask based on
the solar spectrum \citep{chiron1}.  The dips in the cross-correlation
function (CCF) inform us on the RVs, while their amplitudes and widths
help to estimate  relative fluxes and the projected  axial rotation $V
\sin i$.

\subsection{Space Astrometry}
\label{sec:astro}

The knowledge of parallaxes measured by the Hipparcos \citep{HIP2} and
Gaia \citep{Gaia1,Gaia3}  missions is  critical for  interpretation of
the observations. It allows us to convert the fluxes of individual components
(determined by distributing the combined  flux of an unresolved system
between  its  components  using   flux  ratios  deduced  from  the
differential  speckle  photometry  or CCF  parameters)  into  absolute
magnitudes   and   estimate    masses   via   standard   relations
\citep{Pecaut2013}  or  isochrones  \citep{PARSEC}.  Such  masses  are
called  here  {\em photometric}  to  distinguish  them from 
directly measured (``orbital'') masses.  

It is  well known  that astrometry of  unresolved multiple  systems is
often biased \citep{Holl2023}.  The influence of unresolved subsystems
can be gauged  by the reduced unit  weight error (RUWE) in  Gaia or by
the   proper   motion  anomaly,   PMA   \citep{Brandt2018,Brandt2021}.
Luckily, some  triples have wide companions  with unbiased parallaxes,
furnishing accurate  distances.  Orbits  and photometric  masses offer
additional check  of the  parallaxes; alternatively, the  knowledge of
parallaxes and masses helps with poorly constrained orbits.

When a  close pair  is not  resolved directly,  its semimajor  axis is
estimated  from the  period, mass  sum, and  parallax using  the third
Kepler's  law.  This  helps  to interpret  the  astrometric wobble.  A
similar approach is used to estimate  periods of wide pairs from their
projected  separations, assuming  that they  equal the  semimajor axes
(which is true in the statistical sense).  Such crudely estimated long
periods are  denoted as  $P^*$, and the  actually measured  periods as
$P$.

\subsection{Orbit Calculation}
\label{sec:orbit3}

An  IDL code  {\tt orbit3}  that fits  simultaneously inner  and outer
orbits in  a triple system  to available position  measurements and/or
RVs      has     been      used     \citep{ORBIT3}.\footnote{Codebase:
  \url{http://dx.doi.org/10.5281/zenodo.321854}}    The   method    is
presented in  \citet{TL2017}.  The weights are  inversely proportional
to the squares  of adopted measurement errors which  range from 2\,mas
to   0\farcs05  and   more  \citep[see][for   further  discussion   of
  weighting]{Trip2021,Orbits2024}.

The  observed motion  in a  triple system  can be  represented by  two
Keplerian  orbits,  neglecting  mutual  perturbations.   The  relative
positions  are  measured  between  stars, hence  the  outer  positions
include a  contribution from the  inner pair (wobble).   Its amplitude
equals the inner  semimajor axis multiplied by the wobble  factor $|f| =
q_{\rm  in}/(1+q_{\rm  in})$ when  the  inner  pair is  resolved,  or,
otherwise,   by  the   photocenter   wobble  factor   $|f^*|  =   q_{\rm
  in}/(1+q_{\rm in}) - r_{\rm  in}/(1+r_{\rm in})$, where $q_{\rm in}$
and  $r_{\rm  in}$ are  the  mass  and the  flux  ratio  in the  inner
pair. The  wobble factors are  negative when the subsystem  belongs to
the secondary component.  If the inner subsystem is  not resolved, one
of the two  parameters (semimajor axis $a_{\rm in}$  or wobble factor)
must  be fixed,  and only  the wobble  amplitude ($f a_{\rm  in}$
product) is constrained.

When the coverage of the orbit is partial, the data match a wide range
of  potential solutions.   In such  situations, I  fix one  or several
elements  to the  values  that  match the  estimated  masses or  other
constraints  after fitting  the remaining  elements. Thus,  instead of
exploring the full family of potential  orbits, I select one member of
this   family   that  fits   both   the   data  and   the   additional
constraints. Such tentative orbits  still serve as  references for wobble
and usually give a good idea of the mutual orbit inclination $\Phi$.

\subsection{Description of the Tables}
\label{sec:tables}

\begin{deluxetable*}{ l  c rrr rrr r r r r  r }
\tabletypesize{\scriptsize}
\tablewidth{0pt}
\tablecaption{Orbital Elements \label{tab:orb}}
\tablehead{
\colhead{WDS} &
\colhead{System} &
\colhead{$P$} & 
\colhead{$T  $} &
\colhead{$e$} & 
\colhead{$a$} & 
\colhead{$\Omega_A$} &
\colhead{$\omega_A$} &
\colhead{$i$}  &
\colhead{$K_1$} & 
\colhead{$K_2$} & 
\colhead{$V_0$} &
\colhead{$f$} 
 \\
\colhead{HIP}   & &   
\colhead{(yr)} & 
\colhead{(yr)} &
\colhead{ } & 
\colhead{($''$)} & 
\colhead{(degr)} &
\colhead{(degr)} &
\colhead{(degr)} & 
\colhead{(\kms)} &
\colhead{(\kms)} &
\colhead{(\kms)} &
}
\startdata
01316$-$5322 &  Ba,Bb & 4.915     & 2023.96    &  0.595   & 0.0123      & 13.2    & 303.2    & 100.6   & \ldots & \ldots & \ldots & 1.0  \\
7111        &        & $\pm$0.037 &$\pm$0.13   &$\pm$0.120 & $\pm$0.0015&$\pm$4.0 &$\pm$12.8 &$\pm$3.8 & \ldots & \ldots & \ldots & fixed \\
01316$-$5322 &  A,B  & 291.4     & 1845        &  0.30   & 0.639      & 28.5    & 164.1    & 148.2    & \ldots & \ldots & \ldots & \ldots \\
7111        &        & $\pm$4.8  &$\pm$11      & fixed   & $\pm$0.007 &$\pm$10.8 &$\pm$21.6 &$\pm$4.4  & \ldots & \ldots & \ldots & \ldots \\
02460$-$0457 & Ba,Bb & 37.90     & 2000.41      & 0.570     & 0.141     & 51.0     & 35.8     & 30.3     & \ldots & \ldots & \ldots & $-$0.333  \\
12912       &        & $\pm$0.66 &$\pm$0.67     &$\pm$0.063 & fixed     &$\pm$19.7 &$\pm$21.0 &$\pm$5.4  & \ldots & \ldots & \ldots & $\pm$0.029\\
02460$-$0457 & A,B & 1000        & 1898.4       & 0.334     & 1.551     & 146.3     & 46.0     & 126.9     & \ldots & \ldots & \ldots & \ldots  \\
12912       &        & fixed     &$\pm$41.1     &$\pm$0.099 & fixed     &$\pm$11.3 &$\pm$43.7 &$\pm$1.2   & \ldots & \ldots & \ldots &  \ldots\\
03496$-$0220 & Ba,Bb & 0.68749   & 2015.6334    & 0.4095    & 0.0185    & 55.8    & 343.9    & 99.5    & 20.608 & 23.970 & \ldots & $-$0.241 \\
17895        &       & $\pm$0.00026 &$\pm$0.0012 &$\pm$0.0042 & fixed  &$\pm$6.5  &$\pm$0.9  &$\pm$6.8 & $\pm$0.141 &$\pm$0.144 & \ldots & $\pm$0.028 \\
03496$-$0220 & A,B   & 47.19     & 1989.57     & 0.607     & 0.365      & 67.1    & 285.1    & 122.9   & 5.88     & 4.52      & 6.34 & \ldots \\
17895        &       & $\pm$2.71 &$\pm$1.39    &$\pm$0.068 & $\pm$0.018  &$\pm$2.4 &$\pm$2.8  &$\pm$3.3 &$\pm$1.57 & $\pm$0.75  & $\pm$0.10 & \ldots \\
04218$-$2146  & Aa,Ab & 12.515     & 2025.587   & 0.206     & \ldots   & \ldots  & 328.2       & \ldots  & 4.214      & \ldots & $-$3.665 & \ldots \\
20375         &       & $\pm$0.235 &$\pm$0.060  &$\pm$0.012 & \ldots   & \ldots  &$\pm$2.6     &\ldots   & $\pm$0.078 &\ldots  &$\pm$0.099  & \ldots \\
08391$-$5557 & B,C   & 23.47       & 2005.3     & 0.25     & 0.0728    & 173.7    & 263.0      & 155.0   & \ldots & \ldots & \ldots & $-$0.386 \\   
 42424        &       & $\pm$1.08  &$\pm$1.9    &fixed  & $\pm$0.0030 &$\pm$33.0 &$\pm$18.1   &$\pm$6.2 & \ldots & \ldots & \ldots & $\pm$0.029 \\   
08391$-$5557 & A,B    & 1100     & 1869.1     & 0.437    & 1.070       & 60.8    & 304.6      & 162.4   & \ldots & \ldots & \ldots & \ldots \\   
 42424        &       & fixed   &$\pm$18.1    &$\pm$0.049 &fixed &$\pm$83.9      &$\pm$70.2   &$\pm$9.8 & \ldots & \ldots & \ldots & \ldots \\   
10367+1522    &  B,C  & 8.584   & 2011.080    & 0.3398    & 0.1470      & 86.4   &  55.4      & 23.6    & \ldots & \ldots & \ldots & $-$0.500 \\
\ldots        &      & $\pm$0.019 &$\pm$0.023 &$\pm$0.0035 &$\pm$0.0008 &$\pm$2.7 &$\pm$2.6 &$\pm$1.0  & \ldots & \ldots & \ldots & $\pm$0.006 \\     
10367+1522    &  A,B  & 120.0   & 2030.60    & 0.352    & 1.080      & 94.2   &  201.1      & 22.2    & \ldots & \ldots & \ldots & \ldots \\
\ldots        &         &fixed  &$\pm$0.56  &$\pm$0.024   &$\pm$0.029  &$\pm$4.0 &$\pm$3.6 &$\pm$5.5   & \ldots & \ldots & \ldots & \ldots \\     
14040$-$4437  &  Aa,Ab & 0.06569 & 2025.2692 &  0.6534   & \ldots     & \ldots  & 83.52     & \ldots & 54.81   &  58.55  & \ldots  & \ldots  \\
68717         &        & \ldots  & \ldots    &$\pm$0.0009 & \ldots    & \ldots   &$\pm$0.14 & \ldots & $\pm$0.10 &$\pm$0.12& \ldots &  \ldots \\
14040$-$4437  &  A,B   & 414     & 2288      &  0.620   & 1.007       & 127.4   & 277.4     & 62.8 & 2.30  &  4.20  & $-$3.32  & \ldots  \\
68717         &        & $\pm$28  & $\pm$21  & fixed    & $\pm$0.060  & $\pm$2.4&$\pm$2.3 & $\pm$1.2 & fixed & fixed & $\pm$0.04 &  \ldots \\
15474$-$1054  & Ba,Bb  & 0.3640  & 2018.0863  & 0.1125  &   \ldots     & \ldots  & 220.3     & \ldots & 15.193   & 13.649     & \ldots  & \ldots \\
77349         &        & \ldots &$\pm$0.0025  &$\pm$0.0048 & \ldots   & \ldots   &$\pm$2.5   & \ldots & $\pm$0.097& $\pm$0.102 & \ldots & \ldots \\
15474$-$1054  & A,B    & 8.320  & 2014.972   & 0.4366   &  0.2429     & 95.7  & 49.6      & 116.3   & 7.197     & 6.671      & 2.975  & \ldots \\
77349         &       & $\pm$0.011 &$\pm$0.013  &$\pm$0.0027&$\pm$0.0011&$\pm$0.3 &$\pm$0.4 &$\pm$0.3 & $\pm$0.138 & $\pm$0.135 &$\pm$0.062 & \ldots \\
16161$-$3037  & Aa,Ab & 7.47   & 2019.42     & 0.266    & 0.0344       & 243.9    & 251.6    & 145.0  & \ldots  & \ldots & \ldots & 0.516 \\
79076         &      & $\pm$0.14 &$\pm$0.96  &$\pm$0.090 & $\pm$0.0024 &$\pm$46.7 &$\pm$17.1 & fixed   & \ldots & \ldots & \ldots & $\pm$0.066 \\
16161$-$3037  & A,B  & 166.5   & 2086.3     & 0.141    & 0.323       & 166.6     & 124.0    & 150.1   & \ldots  & \ldots & \ldots & \ldots \\
79076         &     & $\pm$8.0 &$\pm$8.2    &$\pm$0.044 & $\pm$0.029  &$\pm$8.8  &$\pm$11.5 & $\pm$3.6 & \ldots & \ldots & \ldots & \ldots \\
18250$-$0135 &   Aa,Ab & 11.50     & 2022.56  & 0.90   & 0.0196       & 34.6     & 111.1     & 102.5    & \ldots & \ldots & \ldots & 1.0 \\
90253        &         & $\pm$0.12 &$\pm$0.10 & fixed  & $\pm$0.0014  &$\pm$4.2  &$\pm$4.4   &$\pm$2.8  & \ldots & \ldots & \ldots & fixed \\
18250$-$0135 &   A,B  & 381.0     & 1890.6  & 0.578   & 0.678        & 175.7     & 44.5      & 93.3    & \ldots & \ldots & \ldots & \ldots \\
90253        &       & $\pm$26.8 &$\pm$3.9 & $\pm$0.041& $\pm$0.031  &$\pm$0.4   &$\pm$7.1   &$\pm$0.5  & \ldots & \ldots & \ldots & \ldots \\
19540+1518  &  Aa,Ab & 67.7  & 1956.14    & 0.695       & 0.262   & 102.7       & 164.8     & 142.4    &  \ldots & \ldots & \ldots & 0.476 \\
97922       &      &$\pm$0.7 &$\pm$0.58   &$\pm$0.014   & fixed   &$\pm$8.9     &$\pm$8.0    &$\pm$7.6    & \ldots & \ldots & \ldots & $\pm$0.024 \\
19540+1518  &  A,B & 3100  & 1802.3    & 0.411       & 3.77      & 3.8         & 12.5       & 145.6      &  \ldots & \ldots & \ldots & \ldots \\
97922       &      & fixed &$\pm$31.8    &$\pm$0.030 & fixed   1  &$\pm$15.1   &$\pm$12.2   &$\pm$2.7    & \ldots & \ldots & \ldots & \ldots \\
20503$-$7502 & Aa1,Aa2 & 0.04222 & 2025.3182  & 0.0296  & \ldots   & \ldots   & 333.2    & \ldots  & 41.74     & 47.55  & \ldots  & \ldots \\
102855       &   & fixed  & $\pm$0.0002   &$\pm$0.0011  & \ldots   & \ldots   &$\pm$1.5  & \ldots & $\pm$0.08 &$\pm$0.09&\ldots  & \ldots \\
20503$-$7502 & Aa,Ab & 0.3533   & 2025.280  & 0.250    & \ldots    & \ldots   & 305.0    & 50.4   & 3.49      & \ldots  & 15.14  & \ldots \\
102855       &   & fixed  & $\pm$0.013     &$\pm$0.017 & \ldots    & \ldots   &$\pm$15.7  & fixed  & $\pm$0.30 & \ldots  & $\pm$0.22 & \ldots \\
\enddata
\end{deluxetable*}

\begin{deluxetable*}{r l l rrr rr l}    
\tabletypesize{\scriptsize}     
\tablecaption{Position Measurements and Residuals (Fragment)
\label{tab:speckle}          }
\tablewidth{0pt}                                   
\tablehead{                                                                     
\colhead{WDS} & 
\colhead{Syst.} & 
\colhead{Date} & 
\colhead{$\theta$} & 
\colhead{$\rho$} & 
\colhead{$\sigma_\rho$} & 
\colhead{(O$-$C)$_\theta$ } & 
\colhead{(O$-$C)$_\rho$ } &
\colhead{Method\tablenotemark{a}} \\
\colhead{(J2000)} & 
& 
\colhead{(JY)} &
\colhead{(\degr)} &
\colhead{(\arcsec)} &
\colhead{(\arcsec)} &
\colhead{(\degr)} &
\colhead{(\arcsec)} &
}
\startdata
01316$-$5322 & A,B &  1986.8590 &     43.7 &   0.8600 &   0.0500 &     $-$1.0 &   0.0513 &  M \\
01316$-$5322 & A,B &  1990.9202 &     42.4 &   0.8120 &   0.0050 &      0.4 &  $-$0.0070 &  s \\
01316$-$5322 & A,B &  1991.2500 &     41.9 &   0.8170 &   0.010  &     $-$0.0 &   0.0006 &  H \\
01316$-$5322 & A,B &  1991.7119 &     42.3 &   0.8130 &   0.0050 &      0.5 &  $-$0.0003 &  s \\
01316$-$5322 & A,B &  1995.8566 &     41.5 &   0.8130 &   0.0100 &      2.4 &  $-$0.0086 &  s \\
01316$-$5322 & A,B &  2016.0000 &     27.1 &   0.8054 &   0.0020 &     $-$0.0 &  $-$0.0017 & G \\ 
01316$-$5322 & A,B &  2021.8906 &     23.7 &   0.7883 &   0.0020 &        0.1 & $-$0.0026 & S 
\enddata 
\tablenotetext{a}{Methods:
G: Gaia;
H: Hipparcos;
M: visual micrometer measurement;
S: speckle interferometry at SOAR;
s: speckle interferometry at other telescopes; 
V: adaptive optics at VLT.
}
\end{deluxetable*}

\begin{deluxetable*}{r l c rrr l }    
\tabletypesize{\scriptsize}     
\tablecaption{Radial Velocities and Residuals (Fragment)
\label{tab:rv}          }
\tablewidth{0pt}                                   
\tablehead{                                                                     
\colhead{WDS} & 
\colhead{Comp.} & 
\colhead{JD} & 
\colhead{RV} & 
\colhead{$\sigma$} & 
\colhead{(O$-$C)$$ } &
\colhead{Instr.\tablenotemark{a}}
 \\
\colhead{(J2000)}  & & 
\colhead{(JD $-$24,00,000)} &
\multicolumn{3}{c}{(km s$^{-1}$)}  &
}
\startdata
03496$-$0220 & Ba & 57257.8899 &    35.183 & 0.300 &     $-$0.675 & CHI \\
03496$-$0220 & Bb & 57257.8899 & $-$26.605 & 0.300 &     $-$0.558 & CHI \\
03496$-$0220 & Ba & 57276.8025 &    27.717 & 0.300 &      0.226 & CHI \\
03496$-$0220 & Bb & 57276.8025 & $-$16.213 & 0.300 &      0.082 & CHI
\enddata 
\tablenotetext{a}{Instruments:
CHI: CHIRON;
LCO: DuPont echelle. 
}
\end{deluxetable*}

The main results of this  work (orbital elements and observations) are
condensed  in three  tables.   Table~\ref{tab:orb}  lists the  orbital
elements  --- visual,  spectroscopic, or  combined. The  notations are
standard: orbital  period $P$,  epoch of periastron  $T$, eccentricity
$e$,  semimajor axis  $a$  (full  or wobble),  position  angle of  the
ascending  node   $\Omega_A$,  longitude  of   periastron  $\omega_A$,
inclination $i$, RV amplitudes of the primary and secondary, $K_1$ and
$K_2$, the systemic  velocity $V_0$, and the wobble  factor $f$.  When
both  positions and  RVs are  available, the  elements match  both, so
$\omega_A$ is the periastron argument of the primary component and $\Omega_A$
is chosen  to fit the position  angles.  The {\tt orbit3}  code always
considers the  inner subsystem  as the primary  component of  the wide
pair.   When  it  is  actually   the  secondary,  the  outer  elements
$\Omega_A$ and $\omega_A$ returned by the code are changed by 180\degr
and the outer RV amplitudes are swapped.

Individual  positions,  their adopted  errors,  and  residuals to  the
orbits  are  listed  in  Table~\ref{tab:speckle},  available  in  full
electronically.  The subsystems are identified  by their WDS codes and
components. Compared  to the  published SOAR  data, the  positions are
corrected for the small systematics determined in \citet{Tok2022f}; in
a  few cases  the speckle  data  are reprocessed.   The second  column
indicates  the subsystem;  for example,  A,B refers  to the  angle and
separation between A  and unresolved pair B, while A,Ba  refers to the
position of  the resolved  secondary Ba  relative to  A.  The  RVs and
their   residuals   to   orbits   are   listed   in   the   electronic
Table~\ref{tab:rv}.   Larger  errors  are  assigned  to  some  RVs  to
down-weight their impact on the orbit.


\section{HIP 7111 (Quadruple)}
\label{sec:7111}


\begin{figure}
\epsscale{1.0}
\plotone{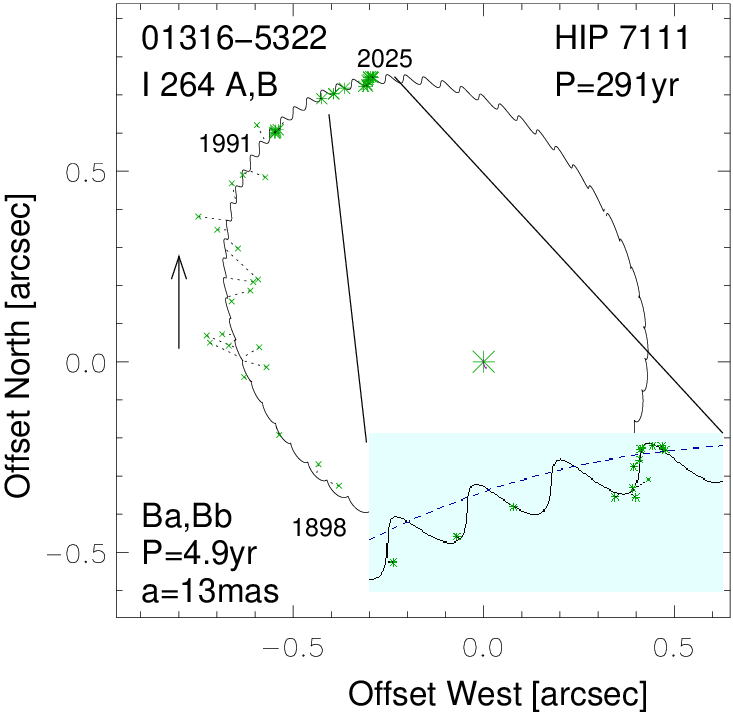}
\caption{Orbits of HIP  7111.  The outer trajectory  with wobble (full
  line),  SOAR   observations  (asterisks),  and   the  center-of-mass
  trajectory  (dashed line)  plotted, with  a zoomed  fragment in  the
  insert.  Less accurate measurements are plotted as crosses.
\label{fig:7111}
}
\end{figure}

This is a visual triple system where  the distant (40\arcsec) companion C has
been identified  by J.~Herschel  in 1836  (HJ 3444), and  the closer
pair  A,B was  discovered  by  R.~Innes in  1898  (I  264).  The  Gaia
astrometry of A and B,  presently at 0\farcs8 separation, is uncertain
(parallaxes  10.61  and  8.83  mas,  mean  parallax  9.72\,mas).   The
distance  to the  system is  known from  the accurate  parallax of  C,
9.677$\pm$0.013\,mas.  Visual orbits of the  A,B pair were computed by
\citet{USN2022} and \citet{Tok2022f},  the latter has a  period of 260
yr. The A,B pair was observed at SOAR frequently as a calibrator until
the wobble in its  motion became obvious (Figure~\ref{fig:7111}).  The
fitted  wobble orbit  has  a period  of  4.9 yr  and  an amplitude  of
12.8\,mas, while the outer period is revised here to 291 yr.  I assume
that the subsystem belongs  to star B because it has  a larger RUWE in
GDR3 (5.3 and 12.4 for A and  B, respectively).  The mass of Ba is 1.2
\msun, the  astrometric orbit gives a  mass of 0.6 \msun  for Bb, and
the full  semimajor axis  of the  Ba,Bb orbit  is 36\,mas.   The outer
orbit  is not  fully  constrained,  so its  eccentricity  is fixed  to
$e_{\rm A,B} =0.3$.   The outer orbit yields a mass  sum of 3.4 \msun,
matching the estimated mass sum of  three stars. Fitting the wobble
leaves weighted residuals  of 2.5 mas; they increase to  7 mas without
wobble.

The ascending  nodes of  both orbits remain  ambiguous, so  the mutual
inclination can be either 50\degr  or 110\degr. The inner eccentricity
of $e_{\rm  Ba,Bb} = 0.6$  provides indirect evidence of  large mutual
inclination and Kozai-Lidov cycles \citep{Naoz2016}.



\section{HIP 12912 (Triple)}
\label{sec:12912}


\begin{figure}
\epsscale{1.0}
\plotone{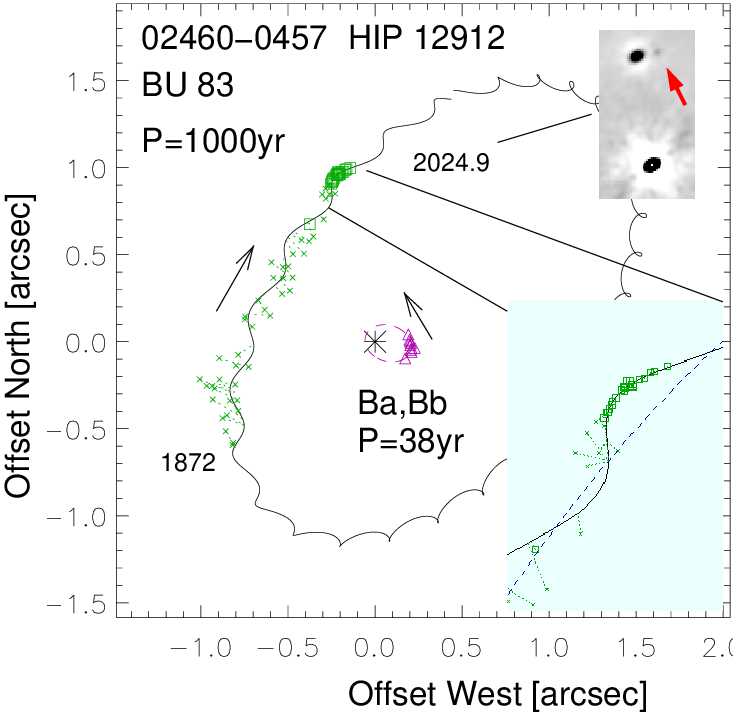}
\caption{Orbits  of HIP  12912. The  inserts  show a  fragment of  the
  speckle autocorrelation image taken in 2024.9 and the zoomed part of
  the  outer orbit  with  the wobble  (full  line), SOAR  observations
  (squares), and  the center-of-mass trajectory (blue  dashed line).
  Less  accurate measurements  are  plotted by  crosses.  The  magenta
  ellipse with triangles shows the inner orbit of Ba,Bb.
\label{fig:12912}
}
\end{figure}

The visual binary  WDS J02460$-$0457 (ADS 2111) has  been resolved for
the  first time  by  S.~W.~Burnham in  1872  (BU~83).  Its  monitoring
during  several  decades  revealed   only  a  slow  motion.   However,
\citet{Dommanget1972}  suspected  a  wave  with a  period  of  36  yr,
presumably caused by an inner  subsystem.  Resolution of the subsystem
by  speckle interferometry  at SOAR  was reported  by \citet{Tok2018}.
Originally, the subsystem  was attributed to the brighter  star A, but
subsequent observations demonstrated that  it belongs to the secondary
star  B.  Here the  updated  analysis  of  this  triple based  on  the
observations till 2024.9  is presented.  The inner period of  38 yr is
confirmed. The  outer orbit  is poorly constrained,  so the  period of
1000 yr is  imposed.  Figure~\ref{fig:12912} shows the  orbits and its
zoomed segment demonstrating  wobble in the motion of  A,B. The wobble
factor  $f=-0.33 \pm  0.03$ implies  the mass  ratio $q_{\rm  Ba,Bb} =
0.49$. The weighted rms residuals  in the outer positions are 4.5\,mas
with wobble and 14\,mas without.

The magnitude difference between Ba and Bb is 3.25$\pm$0.31 mag in the
$I$ filter  (the companion  is not  detected in  the $y$  filter). The
measurements of the inner pair in  2012--2024 cover only a fraction of
its  orbit, while  the  historic  data for  A,B  are  noisy.  In  this
situation, additional constraints on the  masses are helpful. By trial
and error, I  adopted masses of 1.41,  1.05 and 0.58 \msun  for A, Ba,
and Bb, respectively.   The fluxes of such stars derived  from the 2.5
Gyr solar-metallicity isochrone \citep{PARSEC}  match the combined and
differential photometry.  However, the inner mass sum of
1.63 \msun is less than predicted by the unconstrained fit.  To remove
the  discrepancy, the  inner  semimajor axis  is  fixed at  0\farcs141
without increasing the  residuals.  The outer semimajor  axis is also
fixed to match the target mass sum of 3.05 \msun.

Although the  inner and outer  orbits are not well  constrained, their
large mutual  inclination is incontestable, given  the opposite motion
directions.   The  two angles  $\Phi$  computed  from the  orbits  are
124\degr  or   118\degr.   Quasi-orthogonal  orbits  entail   the  von
Zeipel-Kozai-Lidov  cycles  \citep{Naoz2016},  hence the  large  inner
eccentricity $e_{\rm Ba,Bb} = 0.57$ is natural.

The GDR3 provided the 5-parameter astrometric solution for star A with
a small RUWE  of 0.9, and only  the coordinates for star  B (hence the
relative  position  of  A,B  in   GDR3  is  suspect).   No  additional
companions with common PM (CPM) within 2\arcmin are found in GDR3.


\section{HIP 17895 (Triple)}
\label{sec:17895}


\begin{figure*}
\epsscale{0.9}
\plotone{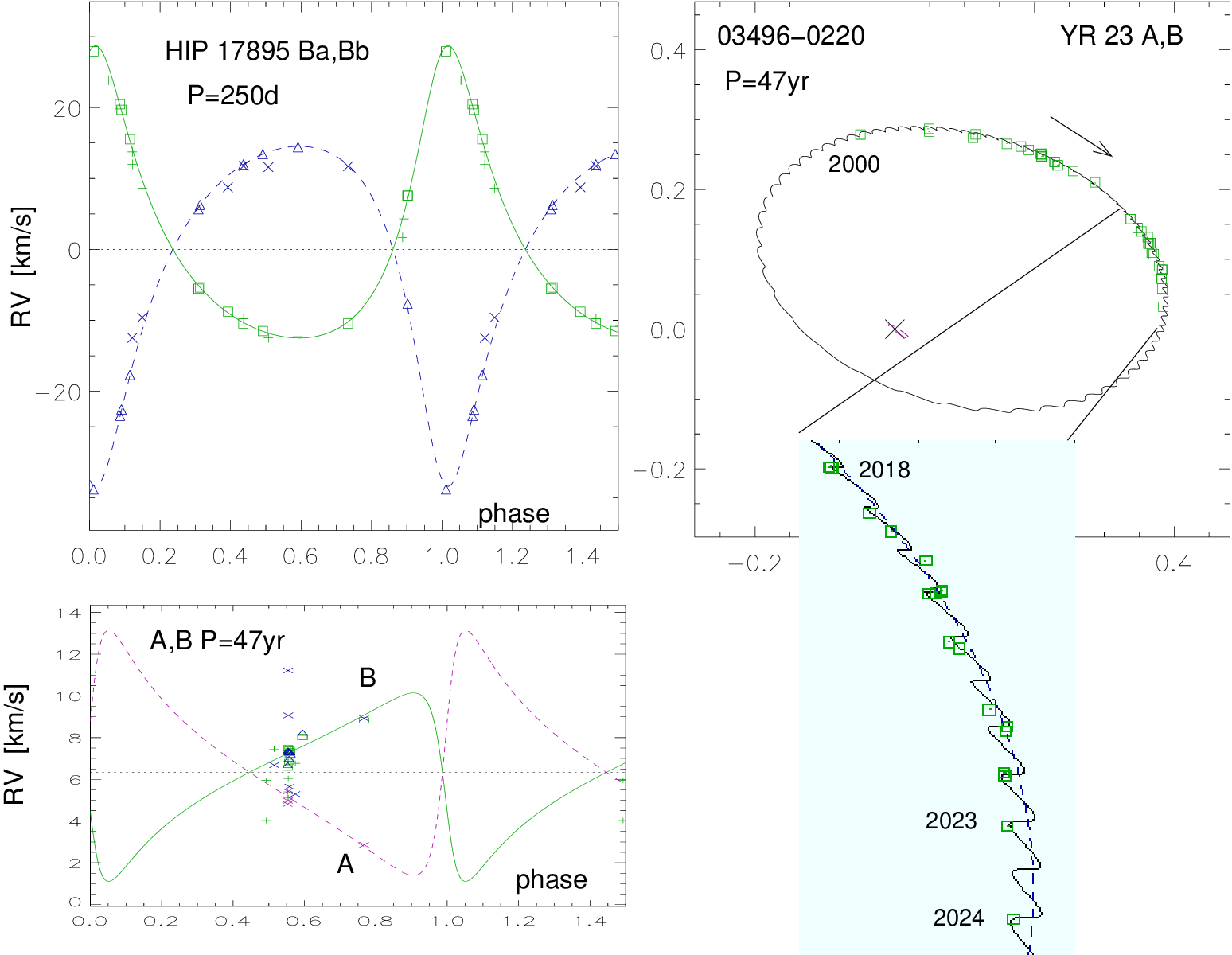}
\caption{Orbits of HIP 17895. The inner and outer RV curves are shown
  on the left (less accurate RVs are plotted by  pluses and
  crosses). In each plot, the motion in other subsystem is
  subtracted. The top-right panel shows the outer orbit on the sky,
  with the latest segment amplified in the insert. 
\label{fig:17895}
}
\end{figure*}

This nearby  solar-type star has been  resolved as a 0\farcs3  pair in
2000 by  \citet{Horch2002}; it received  the designation YR~23  in the
WDS.   In  2015,  its  first, still  preliminary  orbit  was  computed
\citep{RAO2015}, suggesting that the RV  variation  might
be measurable.  Meanwhile, double lines were noted by \citet{N04}. The
triple   nature   of   this   system  was   definitely   revealed   by
\citet{Gorynya2018}  who  presented the  250-day  orbit  of the  inner
subsystem Ba,Bb.  Spectra taken with CHIRON are triple-lined, allowing
to measure the RVs of all three  stars when the lines are not blended.
The dip of A is broadened by axial rotation. 

The semimajor axis of the inner  orbit is 18.5\,mas. As the masses and
fluxes of Ba and Bb are slightly  different, a wobble in the motion of
A,B with an amplitude of 4\,mas  was expected. The YR~23 pair has been
monitored at  SOAR for detecting  this wobble and improving  the outer
orbit; the  results are  presented here  (Figure~\ref{fig:17895}). The
CHIRON spectrum taken in 2025.76  further constrains the RV amplitudes
in  the  outer  orbit.   I  do not  include  some  published  RVs  from
\citet{Gorynya2018} in Table~\ref{tab:rv}.

The inner and outer orbits were fitted jointly. The inner subsystem is
not resolved, so its axis was fixed to the computed value (18.5\,mas).
The wobble factor $f^* = -0.24  \pm 0.03$ corresponds to the amplitude
of  4.4\,mas.  The  weighted  rms  residuals of  the  speckle data  are
1.3\,mas.  Without  wobble, the residuals  increase to 3\,mas,  so the
astrometric effect of  the inner subsystem is  detected reliably.  The
inner inclination is $i_{\rm Ba,Bb} = 99\fdg5 \pm 6\fdg8$; it leads to
the masses of  0.98 and 0.84 \msun for Ba  and Bb, respectively.  With
the photometric  mass of  A, 1.14  \msun, the outer  mass sum  is 2.96
\msun.   The outer  orbit  and the  Hipparcos  parallax of  19.63\,mas
correspond  to the  mass sum  of 2.89  \msun, in  excellent agreement.
However, the  RV amplitudes in  the outer  orbit are not  yet measured
reliably; they  give an outer  mass sum of  only 1.7 \msun.   The next
periastron  in 2036  gives an  opportunity  to measure  better the  RV
amplitudes and other outer elements.

Both  inner and  outer  orbits are  relatively  eccentric.  The  known
orientation of the inner orbit and the correct outer node yield a mutual
inclination of 26\degr, while the period ratio is $\sim$69.


\section{HIP 20375 (Triple)}
\label{sec:20375}


\begin{figure}[ht]
\epsscale{1.1}
\plotone{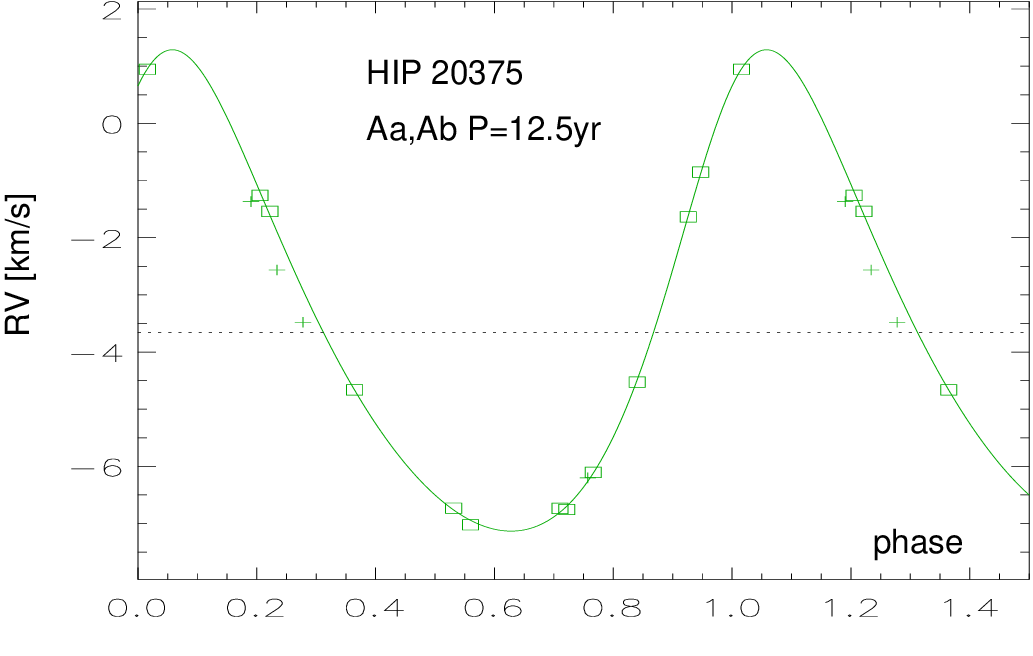}
\caption{Spectroscopic orbit of HIP 20375. Three crosses denote RVs computed
  from the Gaia quadratic trend.
\label{fig:20375} 
}
\end{figure}

The solar-type  (G0V) star HIP 20375  (HD 27723) has been  known as an
unresolved  binary  for   a  long  time,  based   on  its  astrometric
acceleration  \citep{MK05} and  a  slightly  variable RV  \citep{N04}.
Gaia DR3  detected both effects, providing  a 9-parameter acceleration
solution  and a  quadratic RV  trend in  the NSS  catalog \citep{NSS}.
Despite being relatively  bright ($V = 7.54$ mag), this  star does not
have historic  RVs useful for this  study (the data of  Nordstr\"om et
al.  are inaccessible).  The  faint (photometric mass $\sim$0.1 \msun)
CPM companion B at 76\farcs3 makes this system triple. 

This star  has been placed on  the CHIRON program in  2015.6. These 12
RVs covering a time  span of 10.1 yr show a slow  variation. A clue to
the orbital  period is  provided by  the fact that  the PMA  values in
Hipparcos and GDR3  are similar \citep{Brandt2021}.  So,  the 24.75 yr
interval between these missions likely equals one or two periods.  The
RV data were  approximated initially by an orbit with  a fixed 12.4 yr
period; the final  orbit in Table~\ref{tab:orb} has  a similar period,
although its  coverage is not yet  complete. The CHIRON RVs  cover the
same time  as the full Gaia  mission (which will be  released in DR5),
but are more  accurate. The three RVs computed from  the Gaia RV trend
solution show a systematic offset of  $-$0.6 \kms; they are plotted by
crosses  in  Figure~\ref{fig:20375} and  are  not  used in  the  orbit
fit. Suspiciously, the quadratic term in  the Gaia RV trend formula is
7.6 times less than its error.   The mean RV published by \citet{N04},
$-6.2$  \kms,  matches  the  orbit   approximately  if  it  refers  to
$\sim$1985.

The  CCF  dip is  appreciably  widened,  with an  estimated  projected
rotation of $V  \sin i = 19.4$ \kms.  Nevertheless,  the rms residuals
of CHIRON RVs  are only 0.046 \kms.  The lithium  6707\AA ~line is not
seen in the spectra, but an  X-ray radiation from this system has been
detected.  Assuming a  photometric mass of 1.2 \msun  for Aa (spectral
type F8V  and $V-K=1.28$ mag), the  minimum mass of Ab  is 0.45 \msun.
Therefore, non-detection of its lines in the spectrum is natural.  The
semimajor axis of the Aa,Ab pair  is 0\farcs12, so it is resolvable by
adaptive optics; the  estimated contrast in the $K$  band is $\sim$3.1
mag, while $\Delta I_{\rm Aa,Ab} \sim  4.5$ mag is beyond the limit of
speckle-interferometric detection at close  separations. This star was
unresolved  in  the   $K$  and  $H$  bands  at   Gemini-S  in  2011.85
\citep{Tok2012}, when the separation was 44\,mas (see below).

The two remaining unknown orbital elements  can be adjusted to fit the
PMAs in 1991.25  and 2016.0 reported by  \citet{Brandt2021}; $\Omega =
239\fdg5$ and  $i=65\degr$ give a good  match. The mass of  Ab is then
0.51 \msun (spectral type M1.5V), $q_{\rm Aa,Ab} = 0.43$, and $f=0.30$
agrees nicely  with the ratio of  the orbital speed to  the PMA.  This
indicates that  Ab is a  single star, rather  than a close  pair.  The
predicted position of Aa,Ab in 2026.0 is 233\fdg7 and 0\farcs094.  The
pair will become closer in 2029 and will open up to 0\farcs14 in 2033.


\section{HIP 42424 (Triple)}
\label{sec:42424}


\begin{figure}
\epsscale{1.0}
\plotone{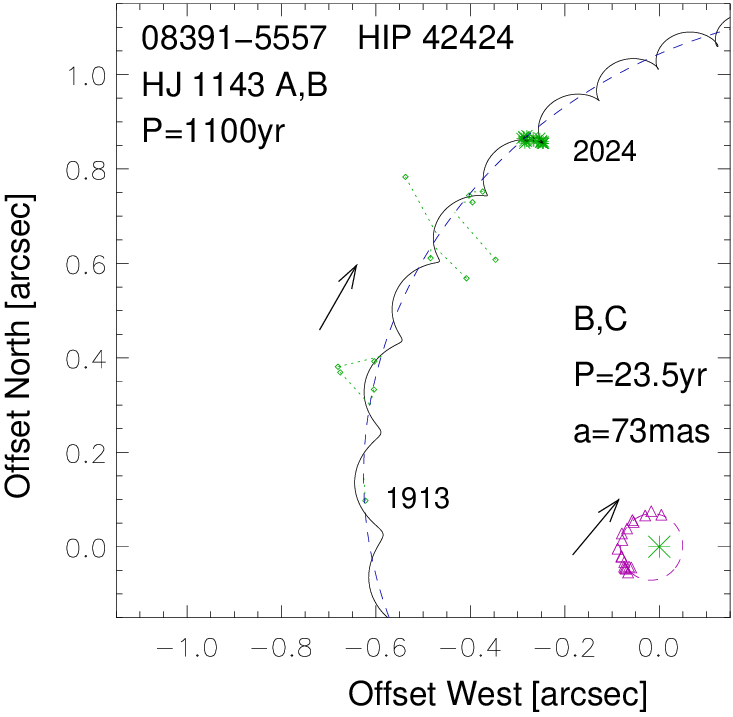}
\caption{Orbits of HIP 42424. Magenta triangles and inner ellipse show
  the orbit of B,C. 
\label{fig:42424}
}
\end{figure}

The  classical visual  binary WDS  J08391$-$5557 (HJ~1443AB)  has been
known since  1913.  Its secondary  component has been resolved  into a
0\farcs09 pair  B,C at SOAR  in 2013. Now  the inner pair  has covered
nearly half  of its orbit,  justifying its preliminary  analysis.  The
outer orbit is  not constrained by the observed arc of 54\degr; yet, the
orbit  of  A,B  with  $P=1447$  yr  and  $e=0.845$  was  published  by
\citet{Izmailov2019}. It corresponds  to a mass sum of  126 \msun with
the GDR3 parallax of 6.22$\pm$0.036 mas  for star A.  A more plausible
(but still tentative)  orbit with $P_{\rm A,B}=1100$~yr and
$e_{\rm  A,B}=0.44$  is  proposed  here.   It  serves,  mainly,  as  a
reference for measuring the wobble.

The two preliminary orbits  are illustrated in Figure~\ref{fig:42424}.
The inner  eccentricity was fixed at  0.25 because the free  fit gives
$e_{\rm B,C}$ with  a large error.  The less certain  outer period and
semimajor axis were  also fixed to values that  match the observations
and the estimated outer mass sum of 4.5 \msun.  The photometric masses
of B  and C, 1.48 and  1.08 \msun, match approximately  the inner mass
sum of 2.9  \msun deduced from the  orbit and the inner  mass ratio of
0.64 inferred from the wobble factor $f = -0.39 \pm 0.03$.  Both inner
and outer pairs are seen nearly  face-on and move clockwise; their two
possible mutual inclinations are between 24\degr and 30\degr.

Gaia DR3 provides astrometric solutions  for star A and the unresolved
pair  B,C.   Their  relative  position,  however,  deviates  from  the
center-of-mass  orbit by  $-10$\degr;  it  is not  used  in the  orbit
fit.  The  inner orbit  can  be  improved  in  the future  if the speckle
monitoring continues for another decade.

\section{DAE 3 (Triple)}
\label{sec:DAE3}


\begin{figure}
\epsscale{1.1} 
\plotone{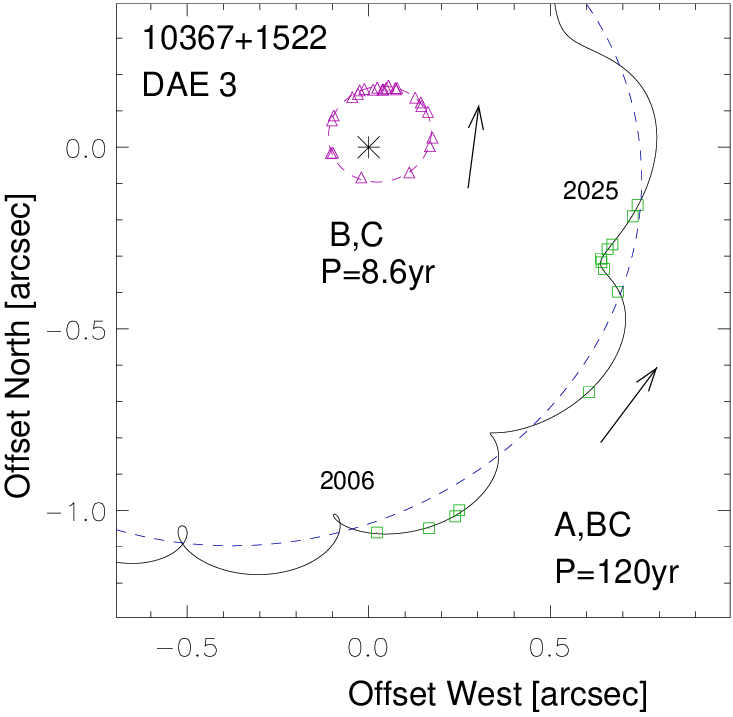}
\caption{Orbits  of DAE~3 (WDS J10367+1522). 
\label{fig:DAE3}
}
\end{figure}

This  low-mass  triple WDS  J10367+1522  (UCAC4  527-051290) has  been
discovered by \citet{Daemgen2007}  in 2006 using adaptive  optics in a
survey of  young M-type  dwarfs.  Lacking  common identifiers  such as
Hipparcos, it  is referred to  here by its WDS  name DAE 3.   The GDR3
parallax of  49.98$\pm$0.09 mas places  the system within  20\,pc from
the Sun.  This triple system has been monitored at SOAR in 2017--2025.
Additional data come from the literature, including accurate positions
from \citet{Calissendorff2022}.  Their  orbit of B,C is  very close to
its present update, which benefits from  six more years
of data. This accurate orbit and  the parallax give the inner mass sum
of 0.344  \msun, so the  two equal-magnitude stars  B and C  are 0.172
\msun  each.   The  wobble  factor  of  $-$0.500$\pm$0.006  confirms  the
equality of the inner masses (Figure~\ref{fig:DAE3}).

The 67\degr arc  of the outer orbit covered by  the measurements gives
only loose  constraints, allowing  periods from $\sim$60  to $\sim$250
years \citep[234\,yr  in][]{Calissendorff2022}; the  unconstrained fit
yields $P_{\rm A,B}  = 120 \pm 34$  yr.  The mass of  star A estimated
from its absolute  magnitude $M_K = 7.06$ mag is  about 0.29 \msun and
corresponds to the spectral type  M3.5V. Therefore, the outer mass sum
should be  around 0.63 \msun.   The outer period  of 120 yr  gives the
outer  mass  sum  of  0.70  \msun,   same  as  the  234  yr  orbit  of
 \citet{Calissendorff2022}. 

The  enforced outer  period  of 120  yr makes  the  two orbits  nearly
coplanar,  $\Phi  =  6\degr  \pm 3\degr$  (the  alternative  angle  is
46\degr), while shorter or longer  periods increase the derived mutual
inclination.  The  planar architecture is common  in low-mass triples,
as well as  equal masses in the inner and  outer pairs (double twins).
In the case  of DAE~3, the inner mass ratio  is indistinguishable from
one, while the outer  mass ratio depends on the mass  of A (0.29 \msun
gives $q_{\rm A,BC} = 0.84$).  Continued monitoring of this remarkable
low-mass  triple system  will  eventually determine  a reliable  outer
orbit,  although  the  progress  will  be  slow.   Historic  positions
measured  on  photographic  plates   could,  potentially,  detect  the
photocenter motion  with the expected amplitude  of $\sim$0\farcs3 and
constrain the  outer period  if the  time base  and accuracy  of these
positions are adequate.


\section{HIP 68717 (Quadruple or Quintuple)}
\label{sec:68717}


\begin{figure*}[ht]
\epsscale{1.0}
\plotone{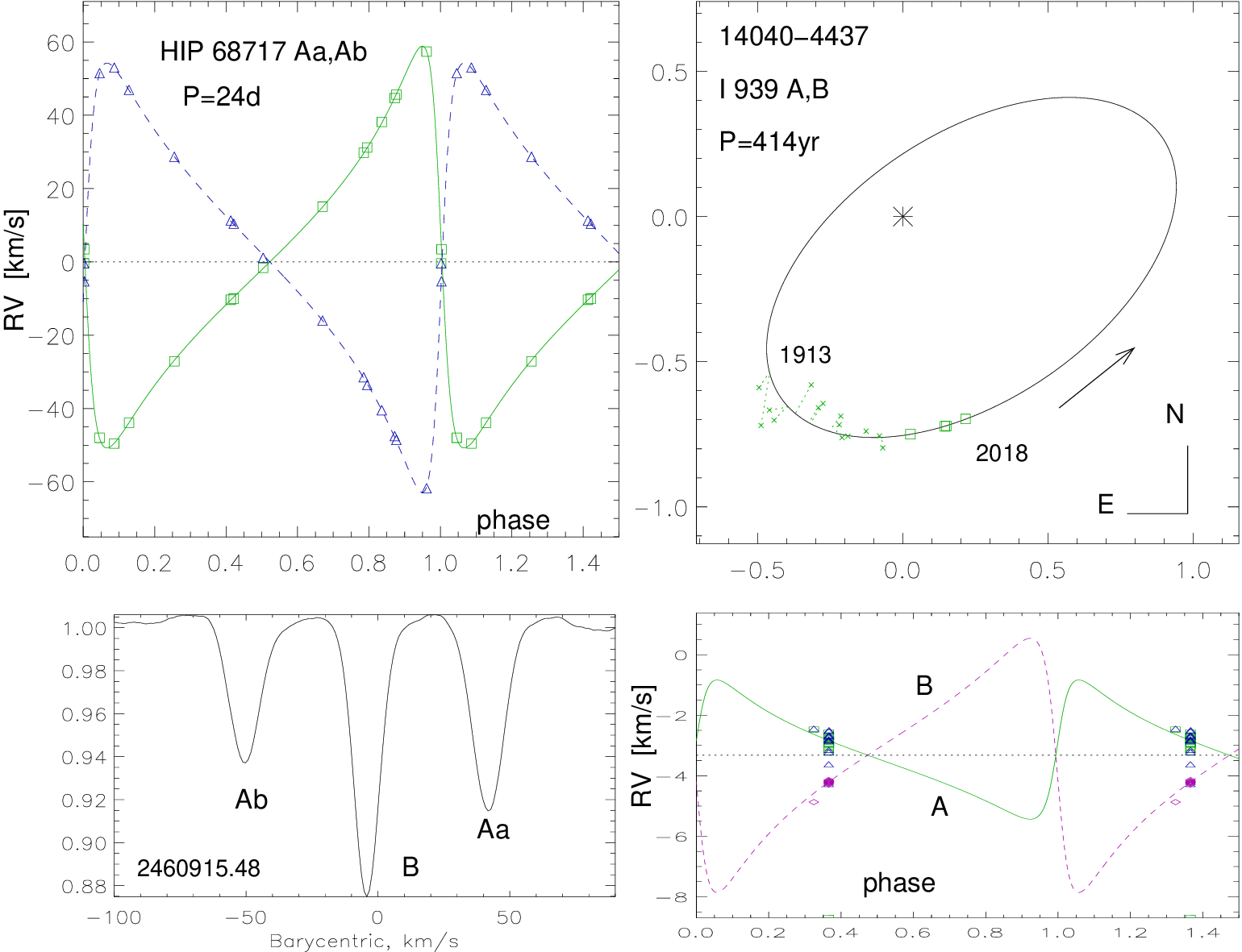}
\caption{Orbits of HIP 68717: the inner RV curve (top left), the outer
  visual orbit, and the outer RV curve.  The triple-lined CCF recorded
  on August 27, 2025 is shown.
\label{fig:68717} 
}
\end{figure*}


The main component of this quadruple system of low-mass stars is known
as HIP~68717,  HD~122613, CPD$-$44~9060,  and WDS  J14040$-$4437.  The
outer 8\arcsec pair  AB,C was noted by  \citet{Hussey1915}, and
A,B has been resolved at 0\farcs6 by  \citet{Innes1911}, although
subsequent micrometer data revealed that this first measure was crude.
More  accurate positions  were measured  by Hipparcos  in 1991  and by
speckle  interferometry at  the Blanco  and SOAR  telescopes in  2008,
2009, and 2018.   The arc of 57\degr covered by  the position measures
does not  constrain the A,B orbit,  and no orbit has  been computed so
far.

Gaia  DR2 measured  a parallax  of 18.95$\pm$0.78  mas for  AB and
3.03$\pm$0.96 mas for  C. However, GDR3 gives only the  positions of A
and B without parallax and a parallax of 12.56$\pm$0.23 mas for C, close
to 11.76$\pm$1.58 mas measured by Hipparcos  for AB. The RVs of AB and
C are  $-3.3$ and $-1.8$  \kms, respectively. The  GDR3 RUWE for  C is
elevated,  9.17,  suggesting  that  it might  contain  an  astrometric
subsystem. The photometry of C ($V=12.43$ mag, $K=9.06$ mag) places it
on the main sequence if its accurate GDR3 parallax is used, while AB is
elevated above  the main  sequence because  it contains  three similar
stars. There is little doubt that  AB and C are physically related and
that the discrepant  astrometry is caused by the  non-single nature of
the sources.

\citet{N04} measured  an RV of $-4.0$  \kms twice over a  time span of
2673 days and  noted double lines.  \citet{LCO} observed  AB and C
with  the DuPont  echelle  in 2008;  they noted  triple  lines in  the
spectrum  of  AB  and  measured  the  RV  of  star  C.  Unfortunately,
monitoring of the triple-lined star AB with CHIRON started only in
2025.

The  lower-left  panel  of   Figure~\ref{fig:68717}  shows  a  typical
triple-lined  CCF computed  from the  CHIRON spectrum.   The strongest
central dip  is identified here  with the  visual secondary B;  its RV
appears constant. The two weaker satellites belong to the double-lined
spectroscopic binary Aa,Ab.  The areas of the dips are 0.63, 0.45, and
0.33 \kms  for B,  Aa, and  Ab, respectively. Their  sum for  Aa+Ab is
larger than for B, indicating a  magnitude difference of 0.23 mag which
agrees with  0.30 mag measured  for A,B by  Hipparcos and $\Delta  I =
0.2$ mag measured at SOAR. 

Frequent observations  of AB  with CHIRON  in April--May  2025 allowed
determination of  the inner spectroscopic orbit  with $P_{\rm Aa,Ab} =  23.99489 \pm
0.00017$ days and $e_{\rm Aa,Ab}  = 0.6534 \pm 0.0009$. The periastron
epoch  is  JD  24660744.5815$\pm$0.0069.   The  period  and  epoch  in
Table~\ref{tab:orb} are expressed in years,  and their tiny errors are
not listed there.

Splitting the combined  flux of AB in proportion to  the measured flux
ratios and  using the parallax of  12.56 mas leads to  the photometric
masses  of   0.99,  0.93,  and   1.04  \msun   for  Aa,  Ab,   and  B,
respectively. The mass  sum of AB, 2.96 \msun, helps  to constrain its
orbit. Another  constraint is provided  by the measured RVs  and their
change  between 2008  and  2025.   The RV  amplitudes  must match  the
estimated masses and the outer orbital elements. By trial and error, I
found elements of  the outer orbit with $P=414$ yr  which satisfy both
the  position  measures  and   the  additional  constraints.  A  fixed
eccentricity of $e_{\rm out} = 0.62$  suffices to match the outer mass
sum.

The final  iteration on both orbits  was made using {\tt  orbit3}.  An
offset of  $-0.5$ \kms  was applied  to the RVs  measured in  2008 (it
accounts for the zero-point), and errors  of 0.5 \kms were assumed for
these RVs. The outer RV amplitudes of  2.3 and 4.2 \kms are imposed to
match the outer mass sum and the mass ratio, and the systemic velocity
of $-3.32  \pm 0.04$ \kms is  obtained from the fit.   The RVs derived
from blended CCFs (by fixing some  dip parameters) are given low weights.
The weighted rms  residuals are 0.15, 0.20, and 0.04  \kms for Aa, Ab,
and B, respectively.   Note that the relative position of  A,B in GDR3
is  discrepant and  not used  in the  orbit fit.   Owing to  the small
magnitude difference between A and B, they were likely swapped in some
Gaia scans, as happened with other similar pairs.

The spectroscopic  mass sum  $M \sin^3  i$ of the  inner pair  is 1.57
\msun, the  photometric mass sum  is 1.91 \msun.  Comparison  of these
numbers  leads  to  the  inclination  $i_{\rm  Aa,Ab}  =  70\degr$  or
110\degr; the outer  inclination is $i_{\rm A,B}  = 62\fdg3$. However,
we do  not know  the direction  of the inner  motion and  the position
angle of  the inner  node, so  the coplanarity  between the  orbits of
Aa,Ab  and  A,B   is  possible,  but  not  proven.   The  large  inner
eccentricity  of 0.65  speaks against  orbit alignment.  The semimajor
axis of the  inner orbit, 2.5 mas, allows its  resolution with VLTI to
measure the relative inclination.

The outer companion C is currently separated from A by 7\farcs86, at a
projected  distance of  6.2 kau.  The photometric  mass of  C is  0.69
\msun, and the estimated period of AB,C  is 8 kyr. Given that the Gaia
astrometry of A and B is not  reliable, I cannot compute the speed of
C relative to AB. Furthermore,  a potential subsystem Ca,Cb suggested
by  the large  RUWE in  GDR3 distorts  the PM.  If this  subsystem
exists, HIP 68717 is a hierarchical quintuple.

The PM of  AB (with the caveat mentioned above)  and its RV correspond
to the  Galactic velocity of  $(U,V,W) = (-4.8, -2.2,  -5.4)$ \kms.
The  CHIRON spectrum  does not  contain the  lithium 6707\AA  ~line or
emissions. The  projected rotation of  three stars estimated  from the
width of CCF dips  is between 6  and 8 \kms. Apparently this system belongs
to the  moderately old disk population.   It is located on  the sky in
the   direction  of   the  Scorpius-Centaurus   association,  but   is
substantially closer to the Sun.


\section{HIP 77349 (Triple)}
\label{sec:77349}


\begin{figure}
\epsscale{1.1} 
\plotone{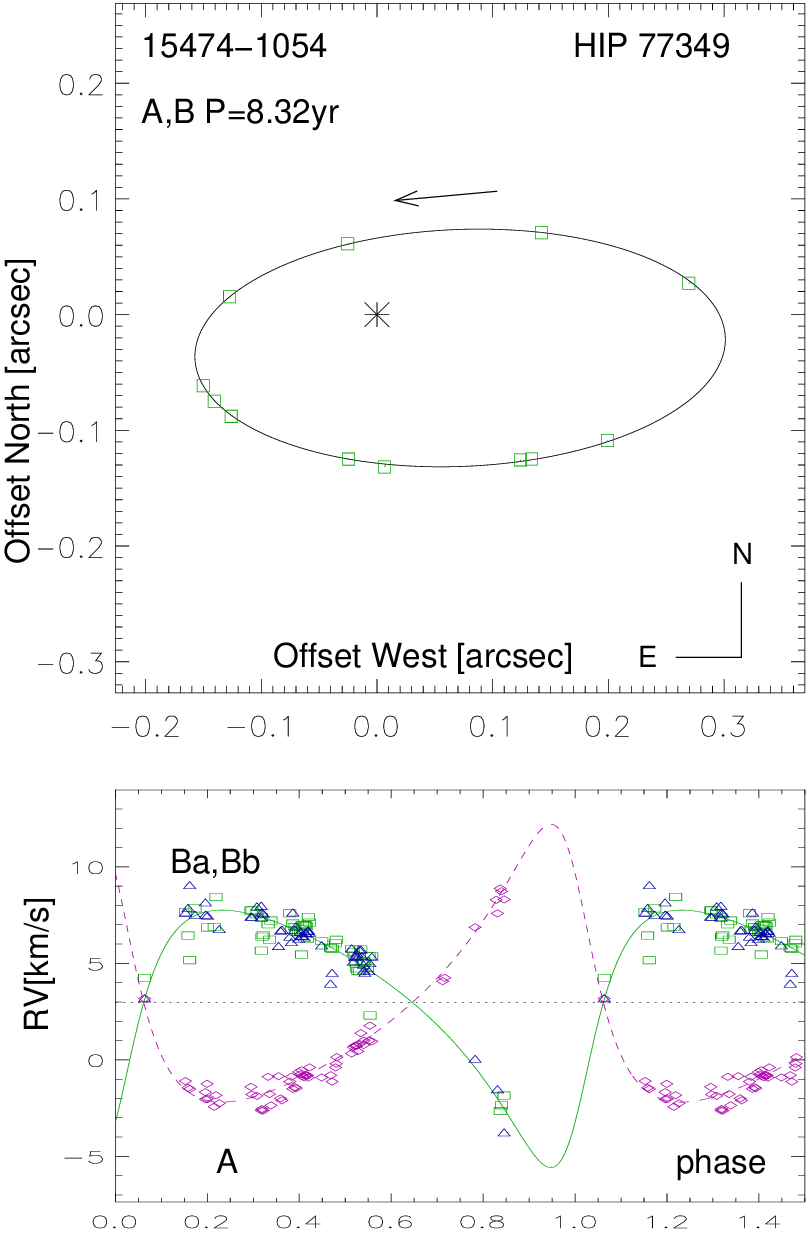}
\caption{Outer orbit  of HIP 77349 on  the sky (top) and  the RV curve
  (bottom), where diamonds, squares, and triangles indicate the RVs of
  A,  Ba,  and Bb,  respectively  (with  subtracted  Ba,Bb
  orbit).
\label{fig:77349}
}
\end{figure}

This nearby  triple system WDS J15474$-$1054  is known  as GJ~3916,  LP~743-31, or
G~152-50;  the spectral  type  is M2.5V.   The GDR3  does  not give  a
5-parameter astrometric  solution, while Gaia DR2  measured a parallax
of 62.52$\pm$0.28  mas. The star  has substantial literature.   It has
been placed on the CARMENES program of RV monitoring and was identified as
a  triple-lined  system   with  periods  of  132.959   and  3082  days by
\citet{Baroch2021},  who also  computed  the  astrometric outer  orbit
based on the Hipparcos data.  The multiplicity inferred from the space
astrometry prompted observations  at SOAR, and the outer  pair has been
resolved in 2019  by \citet{Vrijmoet2022} at 0\farcs23  separation.  Here the
RVs from  \citet{Baroch2021}  are  combined with the  SOAR speckle  data to
derive two very well-constrained orbits.

Given  that  the  RVs  are  published  (they  are  not  duplicated  in
Table~\ref{tab:rv}), I do  not plot here the inner  orbit derived from
the  same  data   (it  is  listed  in   Table~\ref{tab:orb}  only  for
completeness).  The updated  outer orbit using both  positions and RVs
is  shown in  Figure~\ref{fig:77349}.   The accuracy  of 12  position
measurements  covering  5.65 yr  is  excellent,  leaving residuals  of
1.1\,mas.  The  parallax of 62.52\,mas  corresponds to the  outer mass
sum of  0.848 \msun, and  the RV amplitudes  give masses of  0.440 and
0.407  \msun  for  B  and  A, respectively  (sum  0.847  \msun).   The
agreement  indicates that  the Gaia  DR2 parallax  equals the  orbital
parallax. Furthermore,  the known  inner mass  ratio $q_{\rm  Ba,Bb} =
0.90$ gives the directly measured masses of Ba and Bb, 0.208 and 0.232
\msun (Baroch  et al.  designated as  Ba the less massive  star in the
inner pair, and I keep their choice).  The spectroscopic inner mass sum
of 0.323 \msun leads to the  inner inclination of 64\fdg6 or 115\fdg4;
the  latter is  close  to  the outer  inclination  of  $i_{\rm A,B}  =
116\fdg3 \pm  0\fdg3$, suggesting that  the orbits could  be coplanar.
The semimajor  axis of the  inner pair  Ba,Bb is 24.2\,mas,  making it
resolvable at  8 m telescopes (at  SOAR, a marginal elongation  of the
component B  is suspected).   Such a resolution  would yield  a direct
measurement of the mutual inclination.


This relatively compact  triple system composed of M-type  dwarfs is a
double  twin  with  moderate   eccentricities  and,  possibly,  nearly
coplanar  orbits.    Its  resolution   at  SOAR,  together   with  the
triple-lined spectra, offered a unique opportunity to characterize the
architecture almost  completely.  The masses are  measured without any
assumptions.    However,  a   more   elaborate  treatment   (including
estimation  of the  confidence  intervals) and  interpretation of  the
masses and fluxes are outside the scope of this study.

\section{HIP 79706 (Triple)}
\label{sec:79076}


\begin{figure}
\epsscale{1.1} 
\plotone{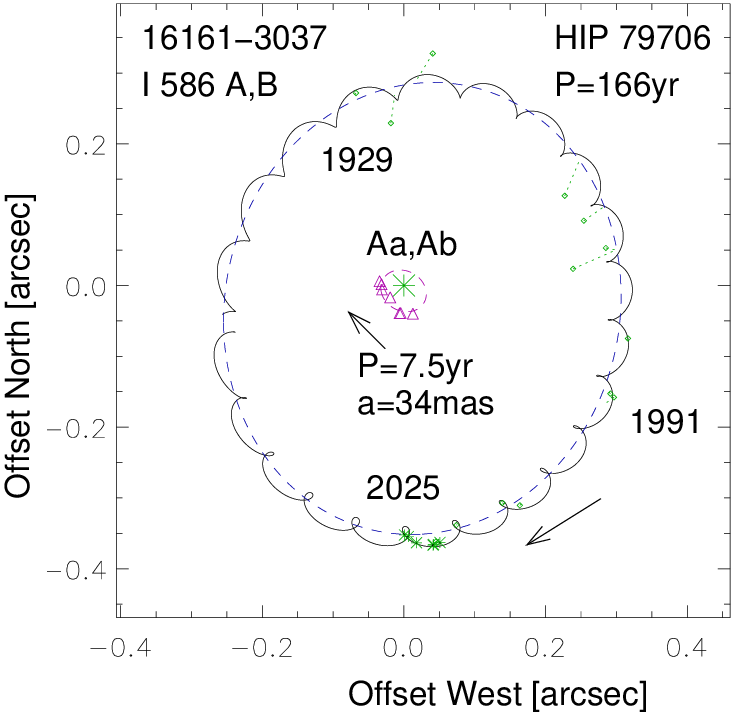}
\caption{Orbits of HIP 79706.
\label{fig:79706}
}
\end{figure}

The visual binary I~1586 AB (WDS J16161$-$3037) has been resolved for the
first time at 0\farcs5 in 1927  by R.~Innes.  The CHARA group measured
3 accurate positions in 1983--2010  which, together with the Hipparcos,
micrometer, and  SOAR positions  (2008--2025) define  the outer  166 yr
orbit rather well. Gaia DR3 did  not resolve the A,B pair and measured
a parallax of 6.20$\pm$0.14\,mas.

The tight (42\,mas)  inner pair Aa,Ab has been first noted  in 2022.4, but,
retrospectively, it can be seen in  the data of 2018.2. Its separation
is  close  to or  below  the  diffraction  limit  of SOAR,  making  the
measurements challenging (reference stars are always needed). By 2025,
the Aa,Ab has  apparently completed one full revolution,  and an orbit
with    $P_{\rm    Aa,Ab}    =    7.5$    yr    is    computed    here
(Figure~\ref{fig:79706}).   While  the  outer  elements  are  measured
without  restrictions,  the  inner  inclination of  $i_{\rm  Aa,Ab}  =
145\degr$ was fixed  (it is within the error bar  of the unconstrained
fit). The  inner pair has a  notable magnitude difference $\Delta  I =
1.53 \pm  0.15$ mag  and $\Delta  y =  1.64 \pm  0.26$ mag.   Yet, the
wobble factor $f = 0.52 \pm  0.07$ suggests equal masses. Adopting the
photometric  masses  of 1.67,  1.21,  and  1.34  for  Aa, Ab,  and  B,
respectively, leads  to the outer and  inner mass sums of  4.2 and 2.9
\msun.  The  parallax   and  the  orbits  give  5.1   and  3.1  \msun.
Considering the uncertainties, the agreement is acceptable.

The mutual  inclination between the  orbits is estimated to  be either
37\degr  or  50\degr, with  a  large  uncertainty.  A  better  angular
resolution and accuracy (hence larger telescopes) are needed to
improve the  inner orbit.   The inner pair  presently closes  down and
will become inaccessible to SOAR in the next few years.


\section{HIP 90253 (Triple)}
\label{sec:90253}


\begin{figure}
\epsscale{1.1} 
\plotone{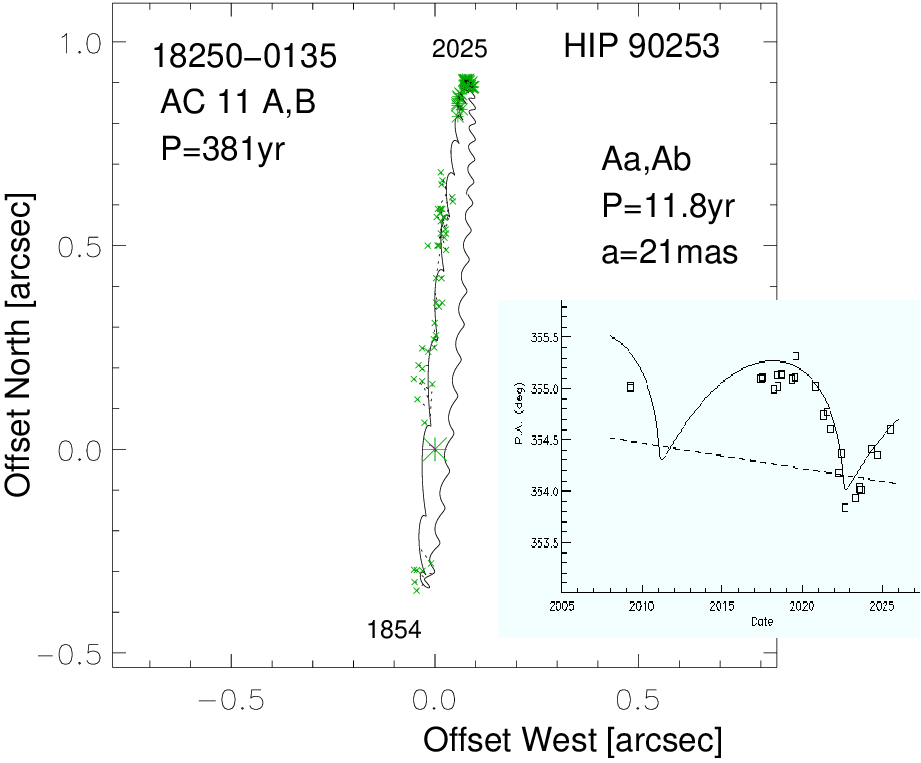}
\caption{Orbit of HIP 90253 A,B with wobble. The insert shows position
  angles  measured at  SOAR vs.  time  (squares), while  the solid  and
  dashed lines depict orbits with and without wobble, respectively.
\label{fig:90253}
}
\end{figure}

The wobble  in the observed motion  of the SOAR calibrator  binary WDS
J18250$-$0135  (HIP  90253,  HR  6898,  AC  11,  ADS~11324)  has  been
suspected for several years.  After 2023, the motion in position angle
reversed    its    ditection,     furnishing    a    decisive    proof
(Figure~\ref{fig:90253}).   A tentative  wobble  orbit  fitted to  the
positions jointly with the outer elements  has a period of 11.8 yr and
a large eccentricity  of 0.90 (fixed).  The inner pair  spends most of
the  time  near  apastron,  offset   from  the  center  of  mass  (the
center-of-mass trajectory  is plotted  by the  dashed line).   The
offset distorts the elements of the  outer orbit, if not accounted for
explicitly.    The  latest   outer  orbit   computed  without   wobble
\citep{Tok2022f} had a period of 434 yr, revised here to 381 yr.  This
pair has been used for  astrometric calibration of speckle instruments
by several teams (e.g CHARA)  and observed frequently for this reason;
here  I  ignore  speckle  data   from  apertures  less  than  2\,m  as
potentially inaccurate.   Gaia DR3 did  not measure the  parallax, the
DR2  parallax is  7.68$\pm$0.17  mas, and  the  Hipparcos parallax  is
7.49$\pm$1.96 mas.

Comparison  between  Gaia  DR2  and  Hipparcos  reveals  a  small  PMA
\citep{Brandt2018}.   This  fact, together  with  the  absence of  the
5-parameter  solution in  GDR3, indicates  that the  subsystem causing
wobble  likely belongs  to the  brighter star  A. The  RV measurements
found in the literature do  not reveal any variability.  Assuming that
the subsystem belongs to star A,  its full axis is 61\,mas, the masses
of Aa and Ab  are 2.3 and 1.6 \msun, respectively (the  F2V star Aa is
slightly  evolved), and  the  mass  of AB  is  5.8  \msun; it  matches
approximately  the outer  mass  sum  derived from  the  orbit and  the
parallax.   The  wobble orbit  is  still  tentative,  it is  not fully
constrained. Two possible mutual inclinations, 139\degr or 40\degr,
however uncertain, exclude mutual alignment and are in harmony with
the large inner eccentricity.



\section{HIP 97922 (Quadruple)}
\label{sec:97922}


\begin{figure}[ht]
\epsscale{1.1}
\plotone{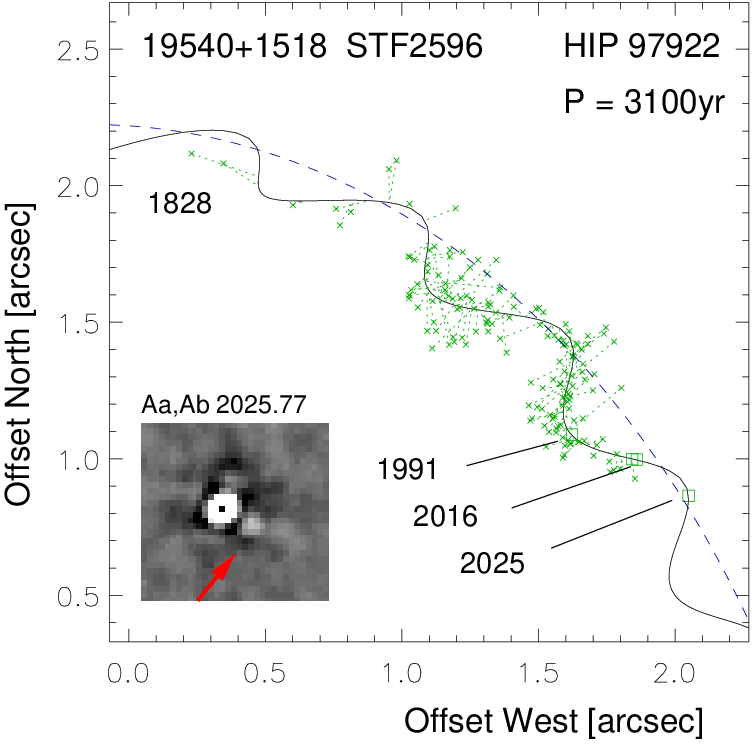}
\caption{Outer  orbit of  HIP  97922.  The  squares  and crosses  show
  accurate and inaccurate measurements,  respectively.  The solid line
  is  an  orbit  with  wobble,  and   the  blue  dashed  line  is  the
  center-of-mass  trajectory.   The  insert   shows  fragment  of  the
  shift-and-add  image of  star A  in the  $I$ filter  with the  inner
  companion Ab on the lower-right side at 0\farcs092 separation.
\label{fig:97922}
}
\end{figure}

The multiple system WDS J19540+1518  (HIP 97922, HD~188328, ADS 13082)
consists of  the outer  pair A,B  discovered by  W.~Struve in  1828 at
2\farcs1 separation (STF 2596) and  the inner 0\farcs09 pair Aa,Ab resolved
in 2025.52 at SOAR and confirmed in 2025.77.  A subsystem
in the component  A is detected independently by the  elevated RUWE of
4.3 in GDR3  and by the PMA \citep{Brandt2021}.  The  parallax of star
B, 11.47 mas (distance modulus 4.70 mag) is adopted here; the parallax
of A, 11.78\,mas, could be slightly biased by the subsystem.  The main
star Aa of spectral type F8III is evolved. 

A  premature  orbit   of  A,B  with  $P_{\rm A,B} = 2971$  yr   was  derived  by
\citet{Izmailov2019} by  a formal fit to  $\sim$190 measures available
in the  WDS database. This orbit  gives an implausible mass  sum of 48
\msun,  while my  estimate  (including the  subsystems)  is 3.8  \msun.
Trying  to compute  a more  realistic outer  orbit, I  found that  the
historic  measures show  a clear  wave with  a period  of $\sim$60
yr. On the other hand, the 0\farcs09 separation of Aa,Ab suggested a period
around 13 yr. Using {\tt orbit3}, I fitted the outer orbit with wobble
that yields  the expected  mass sum  (the data allow  a wide  range of
outer orbits) and, at the same  time, fits the position of Aa,Ab.  The
inner period of 67.8$\pm$0.6 yr and the inner eccentricity of 0.70 are
found; presently Aa,Ab is near the periastron,  opening up slowly.

In 2025 October the main star A  was reobserved at SOAR in the $I$ and
$y$ filters to measure the magnitude difference $\Delta m_{\rm Aa,Ab}$
(2.8 and  3.2 mag,  respectively). Compared  to the  discovery measure
three months earlier, the position  angle of Aa,Ab decreased by 9\degr
and the  separation increased by  2 mas,  in agreement with  the inner
orbit. The RV  of 3.94 \kms measured with CHIRON  in 2025 differs from
10.1 \kms measured by \citet{N04} around 1990.

The  known period  of  Aa,Ab  and the  wobble  amplitude  lead to  the
conclusion that  the new  faint companion  Ab is almost  as massive  as Aa,
hinting that it could be a close pair. If so, eclipses might be found.
Indeed, the TESS \citep{TESS} light curve shows a periodic signal with
$P = 0.27$ day and an amplitude of 3.5\%.  Therefore, this star, known
as a visual binary STF~26596 for two centuries, is in fact a quadruple
system of 3+1 hierarchy.

Fitting both orbits proceeded iteratively. First, the obvious outliers
were  discarded,  and  most  remaining  micrometer  measurements of A,B were
assigned  errors of  0\farcs1.  Speckle  interferometry with  moderate
apertures in 1991--2014  is given errors of 0\farcs05,  while the Gaia
and SOAR relative  positions of A,B are  given errors of 2  and 5 mas,
respectively. As the  outer orbit is not constrained, I  fixed some of
its elements during  iterations, but finally kept only  the period of
3100 yr fixed.  The wobble amplitude of 0\farcs12 is determined securely. 

Figure~\ref{fig:97922}  shows   the  fragment   of  the   outer  orbit
constrained by  the data,  with a wave  in the  historic measurements.
Note  how the  accurate  measurements in  1991,  2015--2016, and  2025
(green squares) deviate from the center-of-mass motion depicted by the
dashed line.  The two orbits  together predict the PM difference B$-$A
of $(-17.6,  -4.0)$ \masyr in  2016, in reasonable agreement  with the
GDR3  PM difference  of $(-18.3,  -2.8)$ \masyr.   The PMA  is $(-9.4,
-1.0)$  \masyr  according  to \citet{Brandt2021}  and  $(-8.0,  -0.5)$
\masyr according to my orbits.

\begin{figure}[ht]
\epsscale{1.1}
\plotone{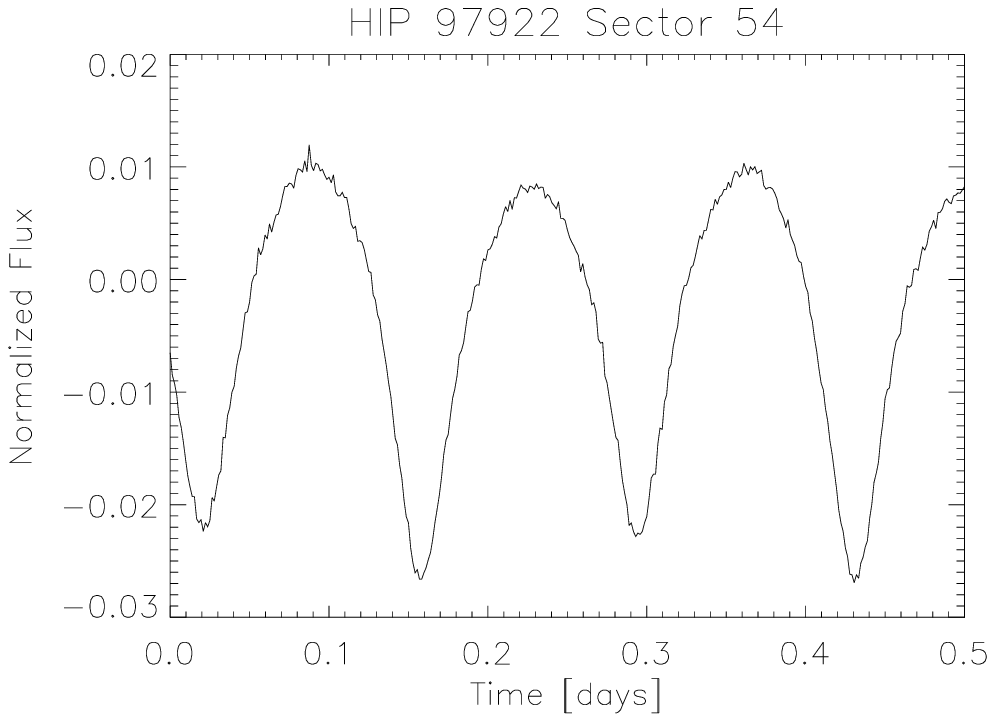}
\caption{Fragment of the TESS light curve of HIP 97922.  Deviations of the
  normalized flux from one are plotted vs. time.
\label{fig:STF2596LC}
}
\end{figure}

This star was monitored by TESS in sector 54 (June 2022); 
the data is available from the MAST archive \citep{MAST}. 
The fragment of the  light curve plotted in Figure~\ref{fig:STF2596LC}
is  typical for  contact eclipsing  binaries.  The  two minima  have a
slightly  different   amplitude,  so   the  true  orbital   period  is
0\fd2730.  Short  periods   are  found  in  other  contact
binaries, e.g.   44~Boo (0\fd268)  and XY Leo  (0\fd284). By  the way,
both these binaries belong to hierarchical systems.

The amplitude of the main minima is 3.5\%. However, the light of Ab is
diluted by  the two brighter stars  Aa and B contributing  to the same
TESS pixel.  The  true eclipse modulation amplitude is  expected to be
$\sim$17  times larger,  about  0.6  mag; it  is  typical for  contact
binaries  of WMa  type.   The light  curve resembles  that  of EK  Com
($P=0\fd2667$)  from  \citet{Liu2023}.  These  authors  determined
physical parameters  of several K-type binaries  with comparably short
periods; for EK  Com they find masses  of 1.02 and 0.28  \msun and the
effective temperature of 5000\,K. These parameters are adopted here as
a  proxy  of  our  binary Ab1,Ab2.   The  standard  relations  between
absolute magnitudes  and periods of contact  binaries \citep{Song2024}
predict  that Ab  has an  absolute $M_G$  magnitude of  $\sim$5.3 mag,
hence $G_{\rm Ab} =  10.0$ mag, 2.9 mag fainter than  Aa ($G_{\rm A} =
7.10$ mag).  This  estimate agrees nicely with  the measured magnitude
difference of $\Delta I_{\rm Aa,Ab} = 2.8$ mag.

\begin{figure}[ht]
\epsscale{1.1}
\plotone{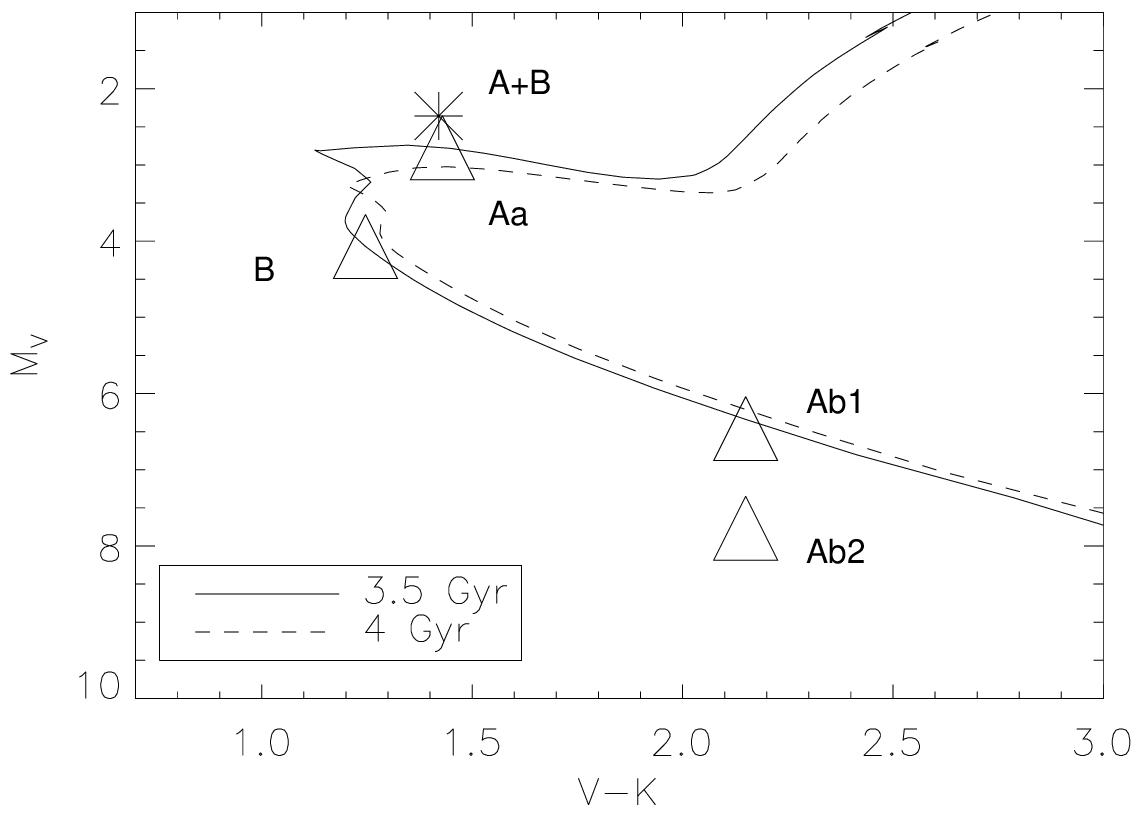}
\caption{Tentative  position of  the components  of HIP  97922 on  the
  color-magnitude  diagram  (triangles).   The  asterisk  denotes  the
  combined  flux  of  four  stars.  The  lines  are  solar-metallicity
  PARSEC isochrones for the ages of 3.5 and 4 Gyr \citep{PARSEC}.
\label{fig:STF2596mod}
}
\end{figure}

\begin{deluxetable}{l c  c l}
\tabletypesize{\scriptsize}     
\tablecaption{Photometry of HIP 97922
\label{tab:STF2592ptm} }  
\tablewidth{0pt}                                   
\tablehead{           
\colhead{Band} & 
\colhead{$m_{\rm A}$} & 
\colhead{$m_{\rm B} - m_{\rm A}$} & 
\colhead{Source} \\
&
\colhead{(mag)} &
\colhead{(mag)} &
}
\startdata
Hp      & 7.372 &  1.149 & Hipparcos \\ 
430nm   & 7.841 &  1.277 & Tycho \\
530nm   & 7.273 &  1.286 & Tycho \\
G       & 7.097 &  1.426 & Gaia DR3 \\
\enddata
\end{deluxetable}

Star Aa is  evolved (F8III) and located well above  the main sequence.
The less evolved star B should be hotter. The resolved photometry of A
and  B  collected  in  Table~\ref{tab:STF2592ptm}  confirms  that  the
magnitude  difference   between  A  and   B  is  smaller   at  shorter
wavelengths.  To  model stellar parameters using  constraints provided
by the photometry, I tried first the  age of 3 Gyr and assigned masses
of 1.445 and 1.18 \msun to Aa and B, respectively. The component Ab is
assumed to  be 3.2 mag fainter  than Aa in  the $V$ band, its  flux is
distributed between  Ab1 and  Ab2 in the  0.7:0.3 proportion,  and the
color  $V-K  =  2.15$  mag  is assigned  to  match  the  Ab  effective
temperature of 5000\,K  (spectral type K2V), appropriate  for a 0\fd27
contact binary.  The model predicts  the combined system magnitudes in
$V$  and $K$  0.25  mag brighter  than observed  (7.06  and 5.64  mag,
respectively).  Increasing the  age to 4 Gyr and  adjusting the masses
yields the system a bit too faint,  and the best match is achieved for
the   age  of   3.5   Gyr.    Figure~\ref{fig:STF2596mod}  shows   the
color-magnitude diagram.   The triangles  are placed on  the isochrone
for the adopted masses  of Aa and B (1.343 and  1.12 \msun), while the
adopted parameters  of Ab1 and  Ab2 are  used for the  eclipsing pair.
The 3.5 Gyr model yields total magnitudes  of 7.15 and 5.73 mag in the
$V$ and $K$  bands, respectively (0.1 mag fainter  than observed), and
the magnitude difference between  A and B of 1.21 mag  in $B$ and 1.37
mag in $I$ (compare to Table~\ref{tab:STF2592ptm}). 

If the eclipsing  binary EK~Com is a  proxy of Ab, its  mass should be
around 1.3 \msun.  The mass sum of Aa,Ab is therefore $\sim$2.6 \msun,
and the mass sum  of A,B is 3.7 \msun.  In the  final iteration on the
orbits,  I fixed  both inner  and outer  semimajor axes  to match  the
estimated mass sums.  The wobble factor  $f = 0.476 \pm 0.026$ implies
the inner mass ratio $q_{\rm Aa,Ab} =  0.9 \pm 0.1$ and the Ab mass of
1.2  \msun.  The  mutual  inclination between  outer and  intermediate
orbits is 52\degr or 45\degr. The  orbit of the eclipsing pair must be
highly inclined,  hence it  is also  misaligned with  the intermediate
orbit  having  $i_{\rm  Aa,Ab}  = 142\degr$.   Kozai-Lidov  cycles  in
misaligned 3+1 quadruples  could be at the origin  of eclipsing pairs,
as suggested by \citet{Hamers2015}.

The adopted masses and orbits  match all observational data reasonably
well.   However,  the true  parameters  may  differ.  Further  speckle
monitoring  of  Aa,Ab  will  eventually  yield  the  model-independent
measurement of the inner mass sum  and the inner mass ratio. The inner
orbit predicts the RV amplitude of 3 \kms for Aa, and the RV variation
is  indeed   detected.   This  3+1  quadruple   system  resembles  WDS
J01350$-$2955 (HIP 7372)  containing a 0\fd5 eclipsing pair  BB Scl as
secondary component in the intermediate  pair with strong wobble.  The
discovery  of the  eclipsing  pair  in HIP  97922  was  driven by  the
unexpectedly large wobble amplitude.


\section{HIP 102855 (Quadruple)}
\label{sec:102855}

\begin{figure}[ht]
\epsscale{1.1}
\plotone{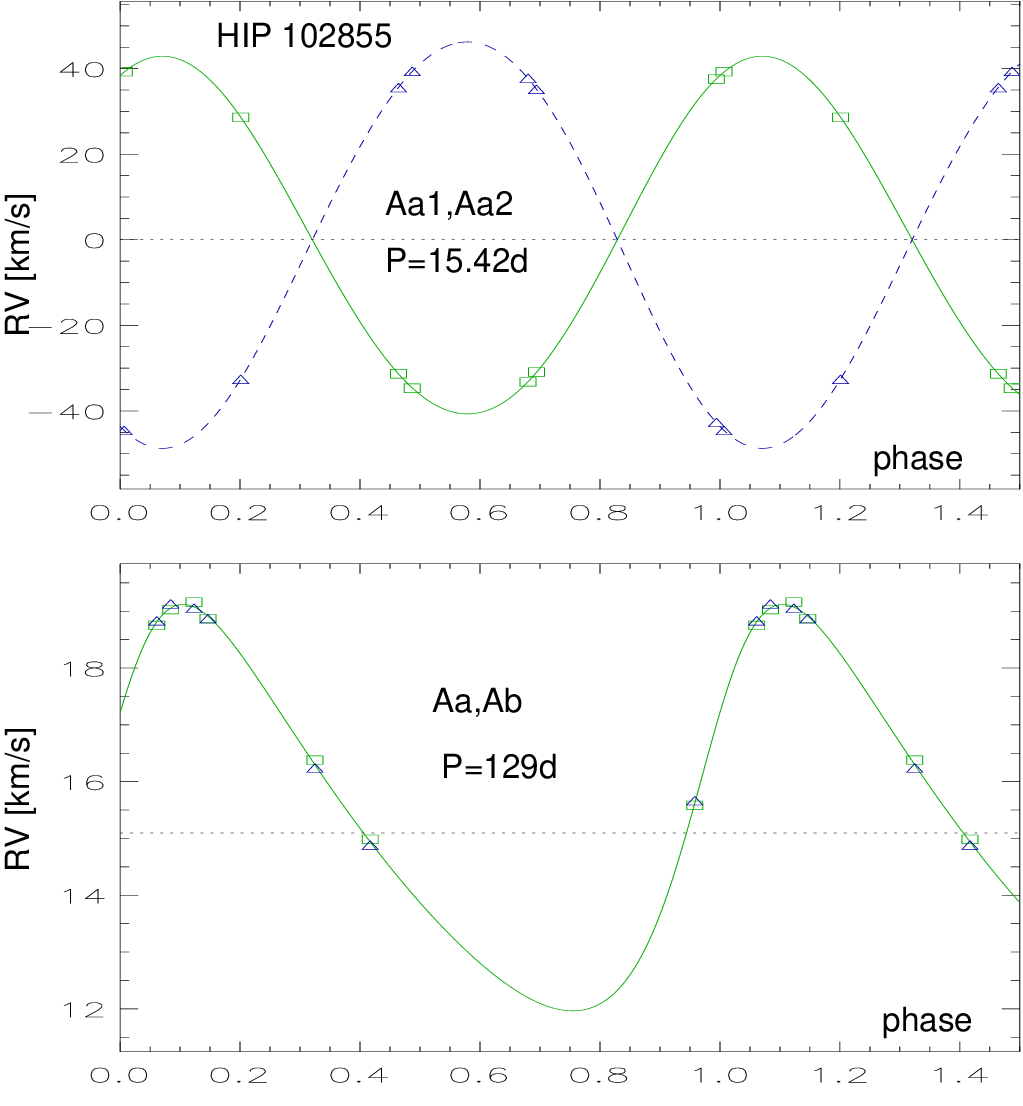}
\caption{Spectroscopic orbits  of HIP  102855 Aa1,Aa2 (top)  and Aa,Ab
  (bottom).  In  each  plot the   contribution  of  other  orbit  is
  subtracted.
\label{fig:102855} 
}
\end{figure}

This inconspicuous 8th magnitude star  (HD 197324, F7V) is featured in
the Gaia  NSS catalog  twice: as  a double-lined  spectroscopic binary
with a 15.4 day  period and as an astrometric binary  with a period of
129 days  and a parallax  of 8.29\,mas.  Some Gaia  astrometric orbits
with  such   periods  are  spurious   \citep[cf.   Figures  16   and  24
  in][]{Holl2023}, so the veracity  of this triple needed confirmation
with CHIRON.  Figure~\ref{fig:102855} shows  a double-Keplerian fit to
the RVs  measured during 2025, where  both periods are fixed  to their
respective NSS values owing to the  small time span of our data.  Both
NSS solutions are  confirmed; for example, the inner  RV amplitudes of
41.74 and  47.55 \kms resemble the  NSS amplitudes of 39.65  and 49.13
\kms, while the inner eccentricity is small (0.030 CHIRON, 0.043 NSS).
The systemic velocity of the inner pair varies with the 129 day period
and an amplitude of 3.49 \kms.   The outer eccentricity of 0.25 fitted
to the RV curve is larger than $e=0.14$ in the NSS astrometric orbit.

The ratio  of CCF dip  areas leads to the  $V$ magnitudes of  8.88 and
9.58  mag  for  Aa1  and   Aa2,  respectively,  corresponding  to  the
photometric masses of 1.31 and 1.06 \msun. The spectroscopic masses $M
\sin^3  i$ are  0.60 and  0.53  \msun, hence  the inner  orbit has  an
inclination of  51\degr or  129\degr. The  outer astrometric  orbit is
inclined at 50\fdg4.   The mass of Ab estimated from  its RV amplitude
and inclination is 0.16 \msun,  while the amplitude of the astrometric
orbit corresponds to 0.19 \msun.  The  contribution of Ab to the total
flux is negligible.

The GDR3  catalog contains a fourth  component B with $G=15.8$  mag at
6\farcs03 separation  and 240\fdg9  angle from A,  with common  PM and
parallax. This companion  has not been noted  previously.  Star B
is also found in the 2MASS  catalog; its estimated mass is 0.38 \msun,
and the period  $P^*$ of A,B deduced from the  projected separation of
727 au is 11 kyr.

The inner compact triple has a  remarkably small period ratio of 8.37.
The  small   inner  eccentricity  and  similar   inclinations  suggest
coplanarity.   The  low-mass  tertiary  Ab apparently  formed  in  the
circumbinary  disc around  Aa1,Aa2  and  had no  chance  to accrete  a
substantial mass while migrating to  its present-day short period. The
growth of Ab could  be prevented by another star B  that formed in the
outskirts  of  the  disk  and preferentially  accreted  the  infalling
gas.


\section{Summary}
\label{sec:sum}

This  work  contributes orbit  determinations  in  a dozen  of  field
stellar hierarchical systems covering a wide range of parameters, from
inner  periods  of  several  days  (spectroscopic)  to  outer  periods
exceeding a millennium,  where a short measured arc gives  only a crude
idea  of the  orbit.   In  such cases,  the  knowledge  of masses  and
distances  helps   to  narrow   down  the  parameter   space.   Mutual
inclinations are determined whenever possible.

The diversity of configurations and parameters is obvious even in this
small  sample  (see   Table~\ref{tab:masses}).   Triple  M-dwarfs  are
represented  by  WDS J10367+1522  (masses  from  0.17 to  0.29  \msun,
periods  8.6  and  $\sim$120  yr)  and a  slightly  more  massive  WDS
J15474$-$1054 (HIP 77349, periods 133 days and 8.3 yr).  These systems
are approximately coplanar (although the  resolution of the inner pair
in  HIP 77349  is  still lacking),  with  moderate eccentricities  and
near-unity mass ratios  in both inner and outer  pairs (double twins).
Such architecture is rather common, e.g. in  HIP 64836 (periods 5 and 30
yr, masses around  0.6 \msun) discussed in the previous  paper of this
series  \citep{Tok2025}  or  the   emblematic  triple  LHS  1070  (WDS
J00247$-$2653) with  periods of 17  and 83 yr. These  low-mass systems
presumably formed in relative isolation  and did not accrete much mass
beyond their parent cores.  Their  planar architecture could be driven
by the angular momentum of the core.

An opposite  case of  misaligned and  eccentric orbits  is exemplified
here  by   HIP  12912  with  counter-rotating   subsystems  and  inner
eccentricity  of 0.57  or  by  HIP 90253  with  inner eccentricity  of
$\sim$0.9.   Characteristically,  stars  in  these  systems  are  more
massive. The quadruple HIP 97922 (discovered by modern observations of
a  well-known,  uninteresting  visual   binary)  is  also  misaligned,
relatively massive (1.34 \msun primary),  and contains a contact inner
binary. The preference of close binaries to be within multiple systems
is   a   well-known   fact  \citep[see   discussion   and   references
  in][]{Mult2021}.  Such  pairs cannot form too  early (otherwise they
would merge).   Dynamics of misaligned  triples and quadruples  is a
potential channel of forming close  binaries via Kozai-Lidov cycles and
tidal friction \citep{KCTF, Fabrycky2007,Naoz2016}.

This  work illustrates  typical problems  of the  multiple-star study.
The main one is the lack  of relevant observations. The WDS archive of
historic micrometer measures  is of invaluable help, but  they are not
available  for newly  discovered hierarchies,  while the  accuracy and
reliability of old  measures are way below  modern standards. Archival
RVs   are   even  less   frequent;   by   an  unfortunate   tradition,
spectroscopists usually  withhold their data until  orbit publication,
even when most  targets are not binary or the  periods are longer than
the time  span of  their program  \citep[e.g.][]{N04}. The  final Gaia
data release 5 will  be limited by the 11 yr  duration of this mission
and by  the angular resolution of  the 1 m Gaia  aperture.  The second
problem is the  complex and intertwined nature  of orbital information
encoded in the  raw data.  While the  automatic pipelines occasionally
fail  in  dealing with  binaries,  the  failure  rate for  triples  is
expected to be even higher. The  example of successful Gaia orbits for
the  inner triple  in  HIP 102855  might be  a  lucky exception.   So,
targeted  ground-based   observations  and  a   scrupulous  individual
treatment of each hierarchy are still  be needed.  This effort will be
rewarded by a deeper understanding of stellar multiplicity.

\begin{acknowledgments} 

The  research  was funded  by  the  NSF's  NOIRLab.   It is  based  on
observations obtained  at the  Southern Astrophysical  Research (SOAR)
telescope, which is a joint  project of the Minist\'erio da Ci\^encia,
Tecnologia e  Inova\c{c}\~{o}es do Brasil (MCTI/LNA),  the US National
Science  Foundation’s NOIRLab,  the  University of  North Carolina  at
Chapel Hill  (UNC), and  Michigan State University  (MSU). It  is alwo
based on observations  at the 1.5 m telescope at  CTIO operated by the
SMARTS  (Small  and  Moderate   Aperture  Research  Telescope  System)
consortium. I thank operators of the 1.5 m telescope for executing
observations of  this program and  the SMARTS team for  scheduling and
pipeline processing.

This work  used the  SIMBAD service operated  by Centre  des Donn\'ees
Stellaires  (Strasbourg, France),  bibliographic  references from  the
Astrophysics Data  System maintained  by SAO/NASA, and  the Washington
Double Star  Catalog maintained at  USNO.  This  work has made  use of
data   from   the   European   Space   Agency   (ESA)   mission   Gaia
(\url{https:\\www.cosmos.esa.int/gaia}),  processed by  the Gaia  Data
Processing        and         Analysis        Consortium        (DPAC,
\url{https:\\www.cosmos.esa.int/web/gaia/dpac/consortium}).    Funding
for the DPAC has been provided by national institutions, in particular
the  institutions participating  in the  Gaia Multilateral  Agreement.
This research has  made use of the data collected  by the TESS mission
funded  by the  NASA Explorer  Program; they  were retrieved  from the
Barbara A. Mikulski Archive for Space Telescopes (MAST).

\end{acknowledgments} 

\facility{CTIO:1.5m, SOAR, Gaia}






\bibliography{triples}
\bibliographystyle{aasjournal}

\end{document}